\documentclass[preprint,eqsecnum,aps]{revtex4}
\usepackage{graphicx}

\newcommand{\AmS}{{\protect\the\textfont2
  A\kern-.1667em\lower.5ex\hbox{M}\kern-.125emS}}

\newcommand{\ee}[2]{ #1 \times 10^{#2} }

\newcommand{\kov}{$\check{\rm C}$erenkov }

\newcommand{\nmne}{\nu_\mu \rightarrow \nu_e}

\newcommand{\nmbp}{\bar \nu_{\mu} p \rightarrow \mu^{+}n}

\newcommand{\nm}{\nu_{\mu}}
\newcommand{\nue}{\nu_{e}}
\newcommand{\nmb}{\bar\nu_{\mu}}
\newcommand{\neb}{\bar\nu_{e}}

\newcommand{\ngs}{^{12}N_{g.s.}}

\newcommand{\necgs}{\  ^{12}C(\nu_{e},e^{-})\ ^{12}N_{g.s.}}

\newcommand{\nec}{\  ^{12}C(\nu_{e},e^{-})\ ^{12}N}

\newcommand{\nmcgs}{\  ^{12}C(\nu_{\mu},\mu^{-})\ ^{12}N_{g.s.}}

\newcommand{\nmbc}{^{12}C(\bar\nu_{\mu},\mu^{+})\ ^{12}B^{*}}

\preprint{LA-UR-01-2390/UCRHEP-E306}
\begin{document}
%
%
%
\vspace{15cm}
\title{ Evidence for Neutrino Oscillations from the \\
Observation of $\bar \nu_e$ Appearance in a $\bar\nu_\mu$ Beam}
\vspace{5cm}
%
%
%
\author{
A. Aguilar,$^5$
L.B. Auerbach,$^8$
R.L. Burman,$^5$ 
D.O. Caldwell,$^3$ 
E.D. Church,$^1$ \\
A.K. Cochran,$^7$\footnote{now at New Brunswick Laboratory, Argonne, IL 60439}
J.B. Donahue,$^5$ 
A.   Fazely,$^7$
G.T. Garvey,$^5$
R.M. Gunasingha,$^7$ \\
R.   Imlay,$^6$ 
W.   C. Louis,$^5$ 
R.   Majkic,$^{8}$
A.   Malik,$^6$
W.   Metcalf,$^6$ 
G.B. Mills,$^5$\\ 
V.   Sandberg,$^5$
D.   Smith,$^4$ 
I.   Stancu,$^1$\footnote{now at University of Alabama, Tuscaloosa, AL 35487}
M.   Sung,$^6$ 
R.   Tayloe,$^5$\footnote{now at Indiana University, Bloomington, IN 47405}\\ 
G.J. VanDalen,$^1$ 
W.   Vernon,$^2$
N.   Wadia,$^6$ 
D.H. White,$^5$ 
S.   Yellin$^3$ \\
(LSND Collaboration)\\
\mbox{}\\
}
\affiliation{$^1$ University of California, Riverside, CA 92521}
\affiliation{$^2$ University of California, San Diego, CA 92093}
\affiliation{$^3$ University of California, Santa Barbara, CA 93106}
\affiliation{$^4$ Embry Riddle Aeronautical University, Prescott, AZ 86301}
\affiliation{$^5$ Los Alamos National Laboratory, Los Alamos, NM 87545}
\affiliation{$^6$ Louisiana State University, Baton Rouge, LA 70803}
\affiliation{$^7$ Southern University, Baton Rouge, LA 70813}
\affiliation{$^8$ Temple University, Philadelphia, PA 19122}
%
%
%
%
\date{\today}
\newpage
%
%
%
\begin{abstract}
A search for $\bar\nu_\mu \to \bar\nu_e$ oscillations was conducted by
the Liquid Scintillator Neutrino Detector at the Los Alamos Neutron
Science Center using $\bar\nu_\mu$ from $\mu^+$ decay at rest.  A
total excess of $87.9 \pm 22.4 \pm 6.0$ events consistent with $\bar
\nu_e p \rightarrow e^+ n$ scattering was observed above the expected
background.  This excess corresponds to an oscillation probability of
$(0.264 \pm 0.067 \pm 0.045)\%$, which is consistent with an earlier
analysis.  In conjunction with other known limits on neutrino
oscillations, the LSND data suggest that neutrino oscillations occur
in the $0.2-10$ eV$^2$/c$^4$ $\Delta m^2$ range, indicating a neutrino
mass greater than 0.4 eV/c$^2$.
\end{abstract}
\maketitle
%
%
%
\section{Introduction}
A neutrino produced in a weak decay is always from a specific family,
$\nu_e$, $\nu_\mu$, or $\nu_\tau$, that is directly associated with
the charged lepton accompanying the decay. When this neutrino is
detected in a charged-current reaction, it manifests its identity by
transforming into the anti-particle of the charged lepton that
accompanied its creation. Lepton family number is then
conserved. However, the result is different if the neutrino changes
from one family to another. For example, if a $\nu_\mu$ changes to a
$\nu_e$, then a $\mu^+$ is made at the neutrino's creation and an
$e^-$ created at its demise, in clear violation of lepton family
number. Such neutrino oscillations are viewed as possible, or even
likely, as the flavor eigenstates ($\nu_e$, $\nu_\mu$, $\nu_\tau$)
need not be neutrino mass eigenstates ($\nu_1$, $\nu_2$, $\nu_3$).  If
the neutrino flavor eigenstates are a linear combination of the mass
eigenstates, the neutrino flavor must change with time because the
phases of the mass eigenstates evolve at different rates. In the case
of two flavor eigenstates ($\nu_a$, $\nu_b$), the probability that
$\nu_a$ will turn into $\nu_b$ is given by
\begin{equation}
P(ab) = \sin^2 ( 2\theta ) \sin^2 \left(1.27 \Delta m^2 {L_\nu \over E_\nu}
\right),
\end{equation}
where $\theta$ is the mixing angle between the mass eigenstates
$\nu_1$ and $\nu_2$, $\Delta m^2$ is the difference in neutrino
eigenstate masses squared, $m_1^2-m_2^2$, in eV$^2$/c$^4$, $L_\nu$ is
the distance traveled by the neutrino in meters, and $E_\nu$ the
neutrino energy in MeV.

A search for neutrino oscillations requires knowledge of the neutrino
source, both with respect to the flavor composition and energy
spectrum of the source. There are two types of searches. The first
seeks to observe a reduction in the expected number of detected
neutrinos of a specific flavor. Characterizing the reduction as
$P(aa)=1- P(ab)$, it can then be explained in terms of neutrino
oscillations.  Such searches are termed {\em disappearance}
experiments. The second method looks for a greater than expected
number of events ascribed to a neutrino flavor that is either absent
or very weakly produced at the neutrino source. These searches are
referred to as {\em appearance} measurements. The results of the
search reported in this paper are of the latter kind. It reports an
excess of events ascribed to electron antineutrinos that is
approximately five times greater than the number of such events
believed to be created at the neutrino source.

Neutrinos are assumed to be massless in the Standard Model, so the
observation of neutrino oscillations would require an extension of the
current version. In addition, as there are $\sim 100$ neutrinos per
cm$^3$ of each neutrino family left over from the initial expansion of
the universe, neutrino masses of even a few eV/c$^2$ would have a
significant effect on the evolving structure of the universe.

The source of neutrinos for the measurement in this report is the
interaction of the intense ($\sim 1$ mA) 798 MeV proton beam at the
Los Alamos Neutron Science Center (LANSCE), which produces a large
number of pions, mostly $\pi^+$.  The $\pi^-$ are mainly absorbed and
only a small fraction decay to $\mu^-$, which in turn are largely
captured.  Thus, the resulting neutrino source is dominantly due to
$\pi^+ \rightarrow \mu^+ \nu_\mu$ and $\mu^+ \rightarrow e^+ \nu_e
\bar \nu_\mu$ decays, most of which decay at rest (DAR).  Such a
source has a paucity of $\bar \nu_e$, and so measurement of the
reaction $\bar \nu_e p \rightarrow e^+ n$, which has a large and well
known cross section, provides a sensitive way to search for $\bar
\nu_\mu \rightarrow \bar \nu_e$ oscillations. Such events are
identified by detection of both the $e^+$ and the 2.2 MeV $\gamma$
from the reaction $n p \rightarrow d \gamma$.  In addition, the
$\nu_e$ flux from $\pi^+$ and $\mu^+$ decay-in-flight (DIF) is very
small, which allows a search for $\nu_\mu \rightarrow \nu_e$
oscillations via the measurement of electrons above the Michel
electron endpoint from the reaction $\nu_e C \rightarrow e^- N$.

The Liquid Scintillation Neutrino Detector (LSND) experiment took data
over six calendar years (1993-1998).  During this period the LANSCE
accelerator operated for 17 months, delivering 28,896 C of protons on
the production target. Using partial samples of the resulting data,
evidence for neutrino oscillations has been published previously
\cite{paper1,bigpaper2,paper3}.  This report presents the final
results on oscillations using all the data, combining the $\bar
\nu_\mu \rightarrow \bar \nu_e$ and $\nu_\mu \rightarrow \nu_e$
searches into a single analysis with common selection criteria, and
employing a new event reconstruction that greatly improves the spatial
resolution. An excess of events consistent with neutrino oscillations
is observed which requires that at least one neutrino have a mass
greater than 0.4 eV/c$^2$.

Neutrino oscillations have also been employed to explain the observed
deficit of solar neutrinos \cite{solar} and the atmospheric neutrino
anomaly \cite{atmos} by $\nu_e$ and $\nu_\mu$ disappearance,
respectively.  The Super Kamiokande atmospheric results \cite{superk}
favor $\nu_\mu \rightarrow \nu_\tau$ and provide compelling evidence
for neutrino oscillations. It is difficult to explain the solar
neutrino deficit, the atmospheric neutrino anomaly, and the LSND
excess of events with only three flavors of neutrinos, so that a
fourth, sterile neutrino has been proposed to explain all of the data
\cite{sterile}.  Neutrino oscillations between active and sterile
neutrinos could have a significant effect on the R process in type II
supernovae \cite{supernovae}.
%
%
%
\section{Neutrino Beam, Detector, and Data Collection}
\subsection{Proton Beam and Targets}

The LSND experiment \cite{bigpaper1} was designed to search for $\bar
\nu_{\mu} \rightarrow \bar \nu_e$ oscillations from $\mu^+$ DAR with
high sensitivity.  A layout of the detector and beam stop is shown in
Fig. \ref{fig:Fig{1}_detector_pic}.  The LANSCE accelerator is an intense
source of low energy neutrinos produced with a proton current of 1 mA
at 798 MeV kinetic energy.  For the 1993-1995 running period the
production target consisted of a 30-cm long water target (20-cm in
1993) followed by a water-cooled Cu beam dump, while for the 1996-1998
running period the production target was reconfigured with the water
target replaced by a close-packed, high-Z target. The muon DAR
neutrino flux with the latter configuration was only 2/3 of the
neutrino flux with the original water target, while the pion DIF
neutrino flux was reduced to 1/2 of the original flux.  The resulting
DAR neutrino fluxes are well understood because almost all detectable
neutrinos arise from $\pi^+$ or $\mu^+$ decay; $\pi^-$ and $\mu^-$
that stop are readily captured in the Fe of the shielding and Cu of
the beam stop \cite{bib:burman}.  The production of kaons or heavier
mesons is negligible at these proton energies.  The $\bar \nu_e$ flux
is calculated to be only $\sim 8 \times 10^{-4}$ as large as the $\bar
\nu_{\mu}$ flux in the $20 < E_{\nu} < 52.8$ MeV energy range, so that
the observation of a $\bar \nu_e$ event rate significantly above the
calculated background would be evidence for $\bar \nu_{\mu}
\rightarrow \bar \nu_e$ oscillations.

For the first three years of data taking, thin carbon targets were in
place in positions A1 and A2 at the experimental area of the LANSCE
accelerator, but dominant pion production occurred at the A6 beam
stop, which accounted for $\sim 98\%$ of the DAR neutrino flux and
$\sim 95\%$ of the DIF neutrino flux.  The A1, A2, and A6 targets were
approximately 135 m, 110 m, and 30 m, respectively, from the center of
the LSND detector.  A6 was essentially the only source of neutrinos
for the last three years of data taking.  Note that in each case there
was a small open space downstream of the primary targets where a few
percent of the pions decay in flight, producing $\nu_\mu$ up to an
energy of 300 MeV.  The neutrino flux was calculated by a program
\cite{bib:burman} using particle production data for thin targets
taken at a number of proton energies and extrapolated to the actual
geometry represented. Fig. \ref{fig:Fig{2}_targets2} shows the layout of the
A6 beam stop as it was configured for the 1993-1995 data taking.
Table \ref{tab:proton_beam} shows the proton beam statistics for each
of the six years of running from 1993 through 1998.

%
\subsection{Neutrino Sources}
Neutrinos arise from both pion and muon decays.  The pion decay modes
are $\pi^+ \rightarrow \mu^+\nu_\mu$, $\pi^+ \rightarrow e^+\nu_e$,
$\pi^- \rightarrow \mu^-\bar\nu_\mu$, and $\pi^- \rightarrow
e^-\bar\nu_e$.  The $\pi^+$ decay occurs both with the pion at rest
($97\%$) and in flight ($3\%$).  The $\pi^-$, however, only decays in
flight as they are totally absorbed on nuclei when they stop.  Helium
represents an anomalous case in which $\pi^-$ decay occurs
occasionally, but this effect is negligible in other nuclei
\cite{bib:piminusdecay}.  Muon decay modes are $\mu^+ \rightarrow
e^+\nu_e\bar\nu_\mu$ and $\mu^- \rightarrow e^-\bar\nu_e\nu_\mu$.
Almost all $\mu^+$ stop before decaying and produce a normal Michel
spectrum for $\nu_e$ and $\bar\nu_\mu$.  The $\mu^-$ are produced
following $\pi^-$ DIF and either decay in orbit or are absorbed in a
nucleus through $\mu^-N \rightarrow \nu_\mu X$, where $E_\nu < 90$
MeV.  The absorption rates are taken from \cite{bib:muabsorp1} and are
shown in Table \ref{tab:muabsorp}.  Each of these production processes
has been included in the flux calculation described below.

\subsection {Production Monte Carlo}

The Production Monte Carlo \cite{bib:burman} simulates the decays of
pions and muons for each of the decay and absorption reactions
described above and for each of the configurations listed in Table
\ref{tab:proton_beam}.  Pion production data using a number of
different proton energies were input, as well as information on the
target materials.  The particles were tracked through the specified
materials and geometries.  For each configuration, the flux and energy
spectrum of neutrinos from each decay channel were obtained for 25
different positions within the detector.  For DAR neutrinos the flux
is isotropic.  The accumulated charge of beam protons was used to
obtain the number of protons on target, and for each year of running
the resulting fluxes and spectra from all configurations were added
together, weighted by the accumulated beam charges.  The program gives
fluxes in terms of the number of neutrinos traversing the detector
region per proton on target per unit of area.

\subsection {Neutrino Fluxes}

Fig.~\ref{fig:Fig{3}_dar_flux} shows the neutrino energy spectra from the
largest DAR sources.  The $\bar\nu_\mu$ flux from $\mu^+$ DAR provides
the neutrinos for the $\bar \nu_\mu \rightarrow \bar \nu_e$
oscillation analysis.  The $\nu_e$ flux from $\mu^+$ DAR provides
events used to verify the DAR neutrino fluxes, as discussed later in
this paper.  The $\bar\nu_e$ flux from $\mu^-$ DAR is a background to
the oscillation signal with an energy spectrum similar to that of
$\nu_e$ from $\mu^+$ decay.

Fig.~\ref{fig:Fig{4}_dif_flux} shows the neutrino energy spectra from various
DIF sources averaged over the detector.  The $\nu_\mu$ flux from
$\pi^+$ DIF provides neutrinos for the $\nu_\mu \rightarrow \nu_e$
oscillation analysis.  The $\nu_e$ flux from $\mu^+$ and $\pi^+$ DIF
is a background for the DIF oscillation analysis. The $\mu^+$ DIF flux
is suppressed due to the long muon lifetime, while the $\pi^+$ DIF
flux is suppressed due to the small $\pi^+ \rightarrow e^+ \nu_e$
branching ratio of $1.2 \times 10^{-4}$.

Calculations of $\mu^+$ DAR fluxes are uncertain at the 7$\%$ level,
while $\pi^\pm$ DIF fluxes and $\mu^-$ DAR fluxes are uncertain to
15$\%$ \cite{bigpaper1}.  Neutrino fluxes for different years are
shown in Table~\ref{tab:fluxes}.

\subsection {Detector}

The LSND detector \cite{bigpaper1} consisted of an approximately
cylindrical tank 8.3 m long by 5.7 m in diameter.  The center of the
detector was 30 m from the A6 neutrino source.  On the inside surface
of the tank, 1220 8-inch Hamamatsu phototubes (PMTs) covered $25\%$ of
the area with photocathode.  The tank was filled with 167 metric tons
of liquid scintillator consisting of mineral oil and 0.031 g/l of
b-PBD.  This low scintillator concentration allows the detection of
both \kov light and scintillation light and yields an attenuation
length of more than 20 m for wavelengths greater than 400 nm
\cite{reeder}.  A typical 45 MeV electron created in the detector
produced a total of $\sim 1500$ photoelectrons, of which $\sim 280$
photoelectrons were in the \kov cone.  PMT time and pulse-height
signals were used to reconstruct the track with an average RMS
position resolution of $\sim 14$ cm, an angular resolution of $\sim
12$ degrees, and an energy resolution of $\sim 7\%$ at the Michel
endpoint of 52.8 MeV.  The \kov cone for relativistic particles and
the time distribution of the light, which is broader for
non-relativistic particles \cite{bigpaper1}, gave excellent separation
between electrons and particles below \kov threshold.  

Cosmic rays were attenuated by roughly 2 kg/cm$^2$ of overburden. The
cosmic ray trigger rate was then reduced from around 10 kHz to an
acceptable level of roughly 50 Hz by an active veto sheild. The veto
shield enclosed the detector on all sides except the bottom, as shown
in Fig. \ref{fig:Fig{1}_detector_pic} by the heavy black line surrounding the
detector.  The main veto shield \cite{veto} consisted of a 15-cm layer
of liquid scintillator in an external tank and 15 cm of lead shot in
an internal tank.  Following the 1993 running, additional counters
were placed over the crack between the endcap veto and the barrel
region veto system, and below the veto shield along the sides. That
reduced cosmic-ray background entering through veto sheild gaps and
the bottom support structure.  The combination of active and passive
shielding tagged cosmic-ray muons that stopped in the lead shot.  A
veto inefficiency $<10^{-5}$ was achieved for incident charged
particles.

\subsection {Data Acquisition}

Digitized time and pulse height of each of the 1220 detector PMTs (and
each of the 292 veto shield PMTs) were recorded when the deposited
energy in the tank exceeded a threshold of 150 hit PMTs ($\sim 4\,{\rm
MeV}$ electron-equivalent energy) with $<4$ veto PMT hits and with no
event with $>5$ veto PMT hits within the previous 15.2 $\mu$s.
Activity in the detector or veto shield during the 51.2 $\mu s$
preceding a primary trigger was also recorded, provided there were
$>17$ detector PMT hits or $>5$ veto PMT hits.  Data were recorded for
1 ms after the primary trigger at a reduced threshold of $21$ PMT hits
(about $0.7 \,{\rm MeV}$) in order to detect the 2.2 MeV $\gamma$ from
neutron capture on free protons, which has a capture time of 186
$\mu$s.  The detector events were recorded without reference to the
beam spill, but the state of the beam was recorded with the
event. Approximately $94\%$ of the recorded events occured between
beam spills, which allowed an accurate measurement and subtraction of
cosmic-ray background surviving the event selection criteria.

As most muons from muon-neutrino induced events do not satisfy the PMT
trigger threshold, these muons were typically past events, while the
electrons from their decay were the primary events.  In contrast,
electrons from electron-neutrino induced events were usually primary
events.  Future events include neutron capture $\gamma$s and
$\beta$-decay electrons and positrons.  Identification of neutrons was
accomplished through the detection of the $2.2\,{\rm MeV}$ $\gamma$
from neutron capture on a free proton.  Nitrogen and boron
ground-state $\beta$-decays occurred after the primary events with
longer lifetimes of 16 and 30 ms, respectively. A given primary event
can have many associated past events and future events.

%
%
%
\section{Neutrino Interactions and Event Simulation}
The neutrino interactions that occured in LSND came from interactions
on carbon, free protons, and electrons in the detector liquid.  All
four possible neutrinos, $\nue$, $\neb$, $\nm$, and $\nmb$, contribute
to neutral-current processes over the entire energy range.
Charged-current cross sections are significantly affected by nuclear
threshold effects. In the case of $\nm$ and $\nmb$ charged-current
interactions, a large amount of the initial neutrino energy goes into
the mass of the final state muon.

Neutrino processes that are observed in LSND are classified into three
categories: standard model leptonic processes (e.g. $\nu e \rightarrow
\nu e$ elastic scattering), inverse $\beta$-decay processes, and
semi-leptonic processes that leave excited or fragmented nuclei in the
final state.  Cross sections in the first category may be calculated
to high accuracy, better than $1\%$, provided that the neutrino energy
is known.  Cross sections for the inverse $\beta$-decay reactions are
inferred from the measured $\beta$-decay lifetimes and are accurate to
the order of a few percent. (The momentum transfers are sufficiently
small that form factor dependences are well characterized.)  The cross
sections for the reactions involving nuclear excited states are much
less certain \cite{Hayes1}.  Models such as the continuum random phase
approximation (CRPA) \cite{kolbe1} often require large corrections in
order to account for ground state wave functions that are too
simplistic.  Fermi Gas models do not reliably take into account
nuclear effects but can be made to produce reasonable agreement in the
quasi-elastic energy region when effective masses are employed
\cite{moniz1,moniz2}.

We use the results of a shell model calculation \cite{Hayes1} for the
$\nec$ DAR processes ($E_\nu < 52.8$ MeV). The shell model calculation
gives a similar energy shape but a lower cross section than the CRPA
calculation \cite{kolbe1}.  A relativistic Fermi Gas model with an
effective mass correction employed to account for nuclear effects is
used for the more energetic DIF neutrino processes.

Two-body neutrino interactions are known accurately from either
measurement or theory.  Those processes are listed in Table
\ref{tab:2B-reactions} with their associated cross section
uncertainty.  They provide the main constraints on neutrino fluxes,
trigger and selection efficiencies, and other neutrino cross sections.
Table \ref{tab:2B-reactions} also lists the neutrino flux sources
constrained by each of these processes.  For example, the $\necgs$ and
the $\nu e \rightarrow \nu e$ elastic reactions primarily constrain
the rate of $\mu^+$ DAR in the target area.  Of all the $^{12}N$
states, only the ground state $\beta$ decays, and the $\nmcgs$
reaction is the best measure of the $\pi^+\rightarrow\nu_\mu \mu^+$
DIF rate in the target area.  Those reactions that contain a final
state $\ngs$ have nuclear matrix elements directly related to well
known nuclear matrix elements, so that the cross sections may be
calculated to an accuracy of a few percent.  The $\nu e \rightarrow
\nu e$ elastic reactions are Standard Model electroweak calculations
and are known to better than a percent from the measured weak mixing
angle, $\sin^2\theta_W$, and the Fermi constant, $G_F$.

Events in the LSND detector were simulated by using the GEANT3.21 code
\cite{geant}, which was modified to track optical photons in addition to
ionizing particles.  Neutrons were tracked and captured on free
protons via the standard MICAP interface to GEANT.  Optical photon yields
from both scintillation and \kov processes were generated and tracked
to the photomultiplier tubes.  A simulation of the photomultiplier
response, analog and digital electronics, and event trigger produced
event data packets, which were a good representation of the LSND
detector response to neutrino events. A large sample of Michel decays
from cosmic rays was used to check the quality of simulated events.
The resulting neutrino event samples are then processed by
the same reconstruction and particle identification software as
the beam data. The reconstructed data are then compared to beam-excess
data for the following analyses.

%
%
%
\section{Data Processing and Event Reconstruction}
A new event reconstruction that improved the position resolution and
the spatial correlation between the $e^+$ and neutron-capture $\gamma$
in the reaction $\bar \nu_e p \rightarrow e^+ n$ was applied to the
entire 1993-1998 data sample.  Different event samples were made
during the new data reduction, and we focus here on the measurement of
electron events, which are relevant to the oscillation search.

The electron selection was applied to the $\sim$4 Terabytes of raw
LSND DLT data tapes, using a minimal set of cuts.  This process
achieved roughly a 40:1 reduction in data size, while maintaining an
$87 \pm 2\%$ efficiency for electron events, independent of electron
energy above 20 MeV.  Events in this new data stream that appeared in
samples from previous LSND analyses were labelled, and a cross-check
for consistency between new and old samples was performed.

Table \ref{tab:electron_selection} shows the electron reduction
criteria and the corresponding efficiencies. First, the visible energy
was required to be greater than 15 MeV in order to eliminate $^{12}B$
$\beta$ decays from cosmic ray $\mu^-$ that stop and capture in the
oil.  Second, the number of associated veto hits was required to be
less than 4. Third, events with a laser calibration tag were
rejected. Fourth, loose electron particle identification criteria were
imposed.  Fifth, the resulting data were subjected to a loose fiducial
volume cut, which required that the reconstructed electron vertex be
inside a volume that was greater than 10 cm from the PMT
surfaces. Finally, cosmic-ray muon events that produced decay
electrons (Michel electrons) as the primary event were removed.  In a
clean sample of cosmic-ray Michel electron events there is a
correlation between the total number of photoelectrons at the muon
time and the online reconstructed distance to the subsequent Michel
electron; as the cosmic muon becomes more energetic, the distance to
the Michel electron grows linearly.  A two dimensional region, or
graphical cut, was imposed to remove these events.

The efficiency for electrons surviving the cuts was determined as
follows.  In an unbiased sample of laser-induced events with their
associated accidental activities, a Monte Carlo (MC) electron event
was inserted in place of the laser ``primary''.  This left a MC
electron event in the midst of the accidental events from the real
laser event.  The MC electrons were generated flat in energy and
uniformly throughout the tank.  Desired accidental properties of the
laser event, {\em e.g.} veto hit count and time to activities, were
preserved when the electron MC event was inserted.  Electron reduction
criteria were then applied and the efficiencies calculated.  The
removal of accidental hits in time with the laser did not
significantly affect the efficiency measurement.

Data were reprocessed with the new event reconstruction in order to
improve the position resolution. The previous event reconstruction was
limited due to the charge response of the 8" PMTs used in LSND
(Hamamatsu R1408). For these PMTs, the single photoelectron output
charge distribution is approximately a broad Gaussian plus an
exponential tail that extends well above the mean of the Gaussian. As
the position and angle fits weighted the hit PMTs by their charge,
this charge tail has the effect of smearing the reconstructed event
positions and angles.  To ameliorate this effect, a new reconstruction
algorithm was developed that weighted the hit PMTs by a ratio of the
predicted charge to the square of the time resolution for that
predicted charge, $Q_{pred}/\sigma_t^2$, and not by their measured
charge, $Q_{tube}$.  (The new reconstruction also has other
improvements, such as the inclusion of timing information in the
$\gamma$ reconstruction.)  This has resulted in an improvement in the
position correlation between the muon and the electron from stopped
muon decay and between the neutron and the $\gamma$ from neutron
capture.  The mean reconstructed distance between the muon and decay
electron improved from 22 cm with the previous reconstruction to 14 cm
with the current reconstruction. For 2.2 MeV $\gamma$ from neutron
capture, the most likely distance was reduced from 74 cm to 55 cm. As
the accidental $\gamma$ rate is proportional to the cube of this
distance, the resulting $\gamma$ reconstruction allows a cut on the
$R_\gamma$ parameter, as described later in section VII, that yields a
factor of two better efficiency with a factor of two reduction in the
rate of accidental $\gamma$s.

%
%
%
\section{Primary Electron Selection}
The primary electron selection is next applied to the reduced data.
The goal of the selection is to reduce the cosmic-ray background to as
low a level as possible, while retaining a high efficiency for
neutrino-induced electron events.  The selection criteria and
corresponding efficiencies are shown in
Table~\ref{tab:electron_selection}.  The energy range $20<E<200$ MeV
is chosen so as to accept both DAR $\bar\nu_\mu \rightarrow \bar\nu_e$
and DIF $\nu_\mu \rightarrow \nu_e$ oscillation candidates.  We
require $20<E_e<60$ MeV for the $\bar \nu_\mu \rightarrow
\bar \nu_e$ oscillation search and $60<E_e<200$ MeV for the
$\nu_\mu \rightarrow \nu_e$ oscillation search.  Below 20 MeV there
are large backgrounds from the $\beta$ decay of $\ ^{12}B$ created by
the capture of stopped cosmic-ray $\mu^-$ on $\ ^{12}C$.  Above 200
MeV the beam-related backgrounds from $\pi^+ \rightarrow e^+\nu_e$ are
large compared to any likely oscillation signal.  Events with a
previous activity within 12 $\mu s$, a future activity within 8 $\mu
s$, or a bottom veto counter hit are rejected in order to eliminate
cosmic-ray muon events.  To further minimize cosmic-ray background, a
tight electron particle identification is applied,
$-1.5<\chi_{tot}^\prime < 0.5$, where the allowed range is chosen by
maximizing the selection efficiency divided by the square root of the
beam-off background with a correlated neutron.  The
$\chi_{tot}^\prime$ parameter depends on the product of the $\chi$
parameters defined in \cite{bigpaper1}.  Briefly, $\chi_r$ and
$\chi_a$ are the quantities minimized for the determination of the
event position and direction, and $\chi_t$ is the fraction of PMT hits
that occur more than 12 ns after the fitted event time. The dependence
of the $\chi$ parameters on energy and position for Michel electrons
was studied, and a correction was developed that made
$\chi_{tot}^\prime$ independent of energy or position.  For the 1993
data only, which had an electronics timing problem resulting in a
broader $\chi_{tot}'$ distribution, $0.3 < \chi_{tot}^{old} < 0.65$
was also required, where $\chi_{tot}^{old}$ is computed like
$\chi_{tot}'$, but with $\chi$ parameters defined in reference
\cite{paper1}.  Additionally, the trigger time is required to occur
between 85 ns and 210 ns in the 500 ns trigger window in order to
reject multiple events, no veto hit is allowed within 30 ns of the
trigger time, and the reconstructed electron vertex is required to be
inside a volume 35 cm from the faces of the photomultiplier
tubes. Finally, the number of associated $\gamma$s with $R_\gamma >10$
($R_\gamma$ is discussed in section VII) is required to be $<2$ ($<1$)
for events $<60$ ($>60$) MeV in order to reject neutron-induced
events, which tend to have many associated $\gamma$s. Neutrons from
$\bar \nu_e p \rightarrow e^+ n$ scattering are too low in energy
($<5$ MeV) to knock out other neutrons; however, higher energy
neutrons ($>20$ MeV) typically knock out 1 or more neutrons.  The
event selection is identical for the DAR and DIF samples except for
the associated $\gamma$ criteria. Note that the event selection is
optimized for electrons in the DAR energy range; however, it was
applied to the DIF energy range for simplicity and in order that a
common selection criteria be used over the entire interval from $20 -
200$ MeV for oscillations from both DAR $\bar \nu_\mu
\rightarrow \bar \nu_e$ and DIF $\nu_\mu \rightarrow \nu_e$. 

In addition to the electron reduction and selection efficiencies,
Table~\ref{tab:electron_selection} also shows the efficiencies due to
the data acquisition (DAQ) and veto deadtime. The total efficiency for
electrons in the fiducial volume with energies in the range $20
<E_e<60$ MeV is $0.42 \pm 0.03$.

%
%
%
\section{Conventional Neutrino Processes}
The neutrino oscillation analysis consists of two steps.  The first
step is to determine the best values for the numbers of events from
standard neutrino processes in a way that minimizes the systematic
uncertainty due to the electron selection.  The second step is to use
those measured neutrino backgrounds as central values in a fit to the
oscillation parameters, allowing the backgrounds to vary around the
central values within their uncertainty. The first step will be
discussed in this section.

The inclusive electron data set provides a common selection for all
neutrino processes important to the oscillation analysis.  Some of
these have well-determined cross sections: the $^{12}N$ ground state
events, $\nu e$ elastic events, and $\bar\nu p$ charged-current
events.  These events serve to constrain the neutrino fluxes and the
selection efficiencies. They also provide important constraints on
uncertain cross sections, such as $^{12}N$ excited state events, where
the nuclear response function is not well known.

Once the primary electron is selected, events are categorized by
whether or not there are associated events in the past or future of
the primary.  This categorization isolates most of the important
reactions.  The simplest event topology has an electron with no
correlated event in the past or future, i.e. inclusive electrons.
Ground state events are selected by searching for $^{12}N$ $\beta$
decay within 70 ms and 70 cm of the primary electron event.  Events
from the process $\bar\nu_e p\rightarrow e^+ n$ have a correlated
$\gamma$ from neutron capture within 1 ms.  Muon neutrino induced
events are selected efficiently because the Michel electron decay of
the muon satisfies the primary electron criteria.  The additional
requirement of a prior event within 10 $\mu s$ efficiently finds the
initial neutrino interaction muon event.  The muon events can have, in
addition, future events from neutron-capture $\gamma$s and nuclear
$\beta$ decays.  A list of the various event categories is shown in
Table \ref{tab:GFcategories}.

A least squares fit was designed to find the best values for the
neutrino fluxes, efficiencies, and cross sections.  It fits those
parameters by minimizing the $\chi^2$ formed from the predicted number
of events in various distributions compared to the observed number of
events.  The distributions are chosen to be sensitive to each of the
parameters in question.  Table \ref{tab:GFparameters} lists the
parameters adjusted in the fit, along with the fitted correction
values, central correction values, and nominal parameter values. The
central correction value is the Gaussian error by which each parameter
was allowed to vary in the fit. The final fitted value for each
parameter is the product of the nominal parameter value and the fitted
correction value.  The agreement between the data and the least
squares fit is good.  The fitted DIF neutrino flux and $\pi^-/\pi^+$
ratio are about one sigma lower than the nominal values; however, the
nominal values are used when estimating the neutrino background to the
oscillation search.  The $\nu_\mu C \rightarrow \mu^- N^*$ cross
section is lower than current theoretical predictions \cite{Hayes1},
\cite{kolbe1} but is in agreement with our earlier measurement
\cite{numuc}.

Fig. \ref{fig:Fig{5}_nuecgs_energy} shows the electron and $\beta$ energy
distributions and the time between the electron and $\beta$, $\Delta
t$, for $^{12}C(\nu_e,e^-)^{12}N_{g.s.}$ scattering events.  The
energy and angular distributions for inclusive electron events are
shown in Fig. \ref{fig:Fig{6}_elec_energy}, where $E_e$ is the electron
energy and $\theta_\nu$ is the angle between the incident neutrino and
outgoing electron directions. Neutrino-electron elastic scattering
events are clearly visible near $\cos \theta_\nu \sim
1$. Fig. \ref{fig:Fig{7}_elec_nuel} shows the angular distribution in more
detail (top plot) as well as the energy distribution (bottom plot) for
the neutrino-electron elastic scattering events with $\cos \theta_\nu
> 0.9$ and with $^{12}C(\nu_e,e^-)^{12}N_{g.s.}$ events removed.
Fig. \ref{fig:Fig{8}_numugs_energy} shows the muon and $\beta$ energy
distributions and the time between the muon and $\beta$ for
$^{12}C(\nu_\mu,\mu^-)^{12}N_{g.s.}$ scattering events. Finally,
Fig. \ref{fig:Fig{9}_numu_energy} shows the Michel electron and muon energy
distributions, the time between the muon and electron, $\Delta t$, and
the distance between the reconstructed electron and muon positions,
$\Delta r$, for $\nu_\mu C \rightarrow \mu^- N$, $\bar \nu_\mu C
\rightarrow \mu^+ B$, and $\bar \nu_\mu p \rightarrow \mu^+ n$
inclusive scattering events.  Cross sections for $\nu C$ scattering
based on a partial data sample have been published previously
\cite{nuec}, \cite{numuc}.  Final cross sections for $\nu e$ elastic
scattering \cite{future_pub1}, $\nu_e C$ scattering
\cite{future_pub2}, and $\nu_\mu C$ scattering \cite{future_pub3} will
be reported elsewhere and are consistent with the nominal parameter
values shown in Table \ref{tab:GFparameters}.

%
%
%
\section{The Decay-at-Rest Oscillation Analysis}
\subsection{Signal and Background Reactions}

The primary oscillation search in LSND is for $\bar \nu_\mu
\rightarrow \bar \nu_e$ oscillations, where the $\bar \nu_\mu$ arise
from $\mu^+$ DAR in the beam stop and the $\bar \nu_e$ are identified
through the reaction $\bar \nu_e p \rightarrow e^+ n$.  This reaction
allows a two-fold signature of a positron with a 52 MeV endpoint and a
correlated 2.2 MeV $\gamma$ from neutron capture on a free
proton. There are only two significant neutrino backgrounds with a
positron/electron and a correlated neutron. The first background is
from $\mu^-$ DAR in the beam stop followed by $\bar \nu_e p
\rightarrow e^+ n$ scattering in the detector. As mentioned earlier,
this background is highly suppressed due to the requirements that a
$\pi^-$ be produced, the $\pi^-$ decays in flight, and the $\mu^-$
decays at rest prior to capture.  The $\bar \nu_e$ flux is calculated
to be only $\sim 8 \times 10^{-4}$ relative to the $\bar \nu_{\mu}$
flux in the $20 < E_{\nu} < 52.8$ MeV energy range.  The second
background is from $\pi^-$ DIF in the beam stop followed by $\bar
\nu_\mu p \rightarrow \mu^+ n$ scattering in the detector.
(Additional contributions are from $\bar \nu_\mu C \rightarrow \mu^+ n
X$ and $\nu_\mu C \rightarrow \mu^- n X$ scattering.) This background
will mimic the oscillation reaction if the $\mu^+$ is sufficiently low
in energy that it is below the threshold of 18 hit PMTs, corresponding
to $E_\mu < 4$ MeV.  Table \ref{tab:sigback} shows the estimated
number of events in the $20<E_e<60$ MeV energy range satisfying the
electron selection criteria for $100\%$ $\bar \nu_\mu \rightarrow \bar
\nu_e$ transmutation and for the two beam-related backgrounds with
neutrons.  Uncertainties in the efficiency, cross section, and $\nu$
flux lead to systematic errors of between $10\%$ and $50\%$ for the signal
and backgrounds discussed below.

The largest beam-related background with a correlated neutron is due
to $\bar\nu_e$ produced in the beam stop by conventional processes.
Such events are identical to the oscillation candidates, and are
identified via the reaction $\bar\nu_e p \rightarrow e^+ n$.  Their
most important source is the DAR of $\mu^-$ in the beam stop.  The
total background due to intrinsic $\bar \nu_e$ in the beam is the
product of neutrino flux ($1.08 \times 10^{11}\bar \nu_e /cm^2$),
average cross section over the entire energy range ($0.72 \times
10^{-40}\,{\rm cm}^2$) \cite{vogel}, the number of free protons in the
fiducial volume ($7.4 \times 10^{30}$), the fraction of events with
$E>20$ MeV (0.806), and the average positron reconstruction efficiency
after cuts (0.42), which gives a total background of $19.5 \pm 3.9$
events before any $\gamma$ selection.  Another possible source of
$\bar\nu_e$, the direct decay of $\pi^- \to e^- \bar\nu_e$, is
negligible, as a consequence of its low branching ratio ($1.2\times
10^{-4}$), the 1/8 ratio of $\pi^-$ to $\pi^+$ in the target, and the
capture of $\pi^-$ in the material of the beam dump.

A related background is due to $\bar \nu_e~^{12} C 
\rightarrow e^+ ~^{11}B~ n$ scattering. The
cross section to the $~^{12}B$ ground state is calculated to be $6.3
\times 10^{-42}$ cm$^2$ \cite{fuku2}, and the cross section to the
$~^{11}B~ n$ final state is estimated to be at least a factor of two
smaller, especially because the first four excited states of $~^{12}B$
are stable against neutron emission.  Therefore, we estimate that this
background is $<2\%$ of the $\bar \nu_e p \rightarrow e^+ n$
background and is negligible.  Furthermore, the maximum positron
energy from this background is 36.1 MeV, so that almost all of the
positrons are below 36 MeV.

The second most important source of beam-related background events
with correlated neutrons is the misidentification of $\bar\nu_{\mu}$
and $\nu_\mu$ charged-current interactions as $\bar\nu_e$ events.
Because of the energy needed to produce a $\mu$, such a
$\bar\nu_{\mu}$ or $\nu_\mu$ must arise from a $\pi$ that decays in
flight. In the tank the $\bar \nu_\mu$ interacts by either
$\bar\nu_{\mu} p \rightarrow \mu^+ n$ or (less often) $\bar\nu_{\mu}
\mbox{C} \rightarrow \mu^+ n X$, followed by $\mu^+ \rightarrow e^+
\nu_e \bar\nu_{\mu}$. The $\nu_\mu$ interacts by $\nu_\mu C
\rightarrow \mu^- n X$.  There are four possible reasons for the
misidentification.  First, the muon can be missed because the $\mu^+$
lifetime is $> 12 \mu$s or the deposited energy is below the 18
phototube threshold for activity triggers.  The latter can occur
either because the muon is too low in energy or is produced behind the
phototube surfaces.  The detector Monte Carlo simulation is used to
show that this threshold corresponds to a $\mu$ kinetic energy,
$T_{\mu}$, of approximately $3\,{\rm MeV}$.  The background rate from
$\bar\nu_{\mu} p \rightarrow \mu^+ n$ is written as the product of the
total $\bar\nu_{\mu}$ flux above threshold ($2.56 \times 10^{11}\,\bar
\nu_{\mu} /{\rm cm}^2$), the average flux-weighted cross section ($4.9
\times 10^{-40}\,{\rm cm}^2$) \cite{vogel}, the fraction of $\mu^+$
having $T_{\mu} < 3\,{\rm MeV}$ or $\tau_\mu >12 \mu$s (0.0258), the
number of free protons in the fiducial volume ($7.4 \times 10^{30}$),
the positron efficiency (0.42), and the fraction of events with
$E>20\,{\rm MeV}$ (0.816), for a background of 8.2 events.  Similar
estimates for the backgrounds from $\bar\nu_{\mu} \mbox{C} \rightarrow
\mu^+ n X$ and $\nu_{\mu} \mbox{C} \rightarrow \mu^- n X$ \cite{kolbe}
add 0.4 and 1.4 events, respectively, for a total of $10.0 \pm 4.6$
events.  It is estimated \cite{kolbe} that about $80\%$ of the $\bar
\nu_{\mu} C \rightarrow \mu^+ X$ and $6\%$ of the $\nu_{\mu} C
\rightarrow \mu^- X$ scattering events will have a recoil neutron.

Second, a $\mu$ above the hit threshold can be missed if a prompt
decay to $e$ caused the muon and electron to be collected in a single
event which is then misidentified as an $e$.  This effect is
considerably suppressed by the electron selection and the requirement
that the reconstructed time be consistent with the triggered event
time.  The detector Monte Carlo simulation shows that this
misidentification only occurs for $\mu^+$ decays within $100\,{\rm
ns}$, decreases with $T_{\mu}$ , and is almost zero above $10\,{\rm
MeV}$. Using the Monte Carlo misidentification probabilities, a
calculation similar to that above implies a background of $0.2 \pm
0.1$ events.

Third, the $\mu^+$ can be lost because it is produced behind the PMT
surface and the electron radiates a hard $\gamma$ that reconstructs
within the fiducial volume. A background of $0.2 \pm 0.1$ events is
estimated from the Monte Carlo simulation.

Fourth, a muon can be missed by trigger inefficiency. After 1994, we
acquired for many online positron triggers complete digitization
information for all veto and detector PMTs over the 6 $\mu$s interval
prior to the positron. Analysis of these data, discussed below, shows
the trigger inefficiency for low-energy muons to be negligible.

There are additional backgrounds from $\bar\nu_e$ produced by $\mu^-
\rightarrow e^- \nu_{\mu} \bar\nu_e$ and $\pi^- \rightarrow e^-
\bar\nu_e$ DIF.  These $\bar\nu_e$ can interact on either $C$ or a
free proton to yield the oscillation signature of a positron and a
recoil neutron.  For $20 < E_e < 60\,{\rm MeV}$, $0.1 \pm 0.1$ events
are estimated.  The reactions $\nu_e ~^{12}C \rightarrow e^- n X$ and
$\nu_e~^{13}C \rightarrow e^- n X$ are negligible ($<0.1$ events) over
the $20<E_e<60$ energy range and cannot occur for $E_e>20$ MeV and
$E_e>36$ MeV, respectively.  Other backgrounds, for example $\nu_{\mu}
C \rightarrow \nu_{\mu} n \gamma X$ with $E_{\gamma} >20$ MeV, $\nu_e
C \rightarrow e^- p X$ followed by $~^{13}C(p,n)~^{13}N$, and $\nu_\mu
C \rightarrow \mu^- X$ followed by $\mu^-$ capture, are also
negligible.

The total background due to pion and muon DIF is $10.5 \pm 4.6$ events
before any $\gamma$ selection. It has a detected energy spectrum which
is very close to that for positrons from $\mu^+$ decay.

A final source of background is neutrons from the target that find
their way into the detector tank. However, a stringent limit on beam
neutron background relative to the cosmic neutron background has been
set by looking for a beam-on minus beam-off excess of neutron events
that pass neutron PID criteria in the 40-180 MeV electron equivalent
range \cite{bigpaper2}.  No excess has been observed, which implies
that the beam-related neutron background is less than $1\%$ of the
total beam-unrelated background and is negligible.

The number of events expected for $100 \%$ $\bar\nu_{\mu} \rightarrow
\bar \nu_e$ transmutation followed by $\bar \nu_e p \rightarrow e^+ n$
scattering (plus a small contribution from $\bar \nu_e C \rightarrow
e^+ B n$ scattering) is $33300 \pm 3300$ events, where the systematic
error arises from uncertainties in the neutrino flux ($7\%$) and $e^+$
efficiency (7$\%$).  This number of events is the product of neutrino
flux ($1.26 \times 10^{14} \nu$/cm$^2$), the average cross section
\cite{vogel} over the entire energy range ($0.95 \times 10^{-40}$
cm$^2$), the average positron reconstruction efficiency (0.42), the
fraction of events with $E>20$ MeV (0.894), and the number of free
protons in the fiducial volume ($7.4 \times 10^{30}$).

\subsection{The Positron Criteria}

The positron/electron selection criteria (LSND is insensitive
to the sign of the charge) for this primary oscillation search is 
described in detail in section V.

\subsection{The Correlated 2.2 MeV $\gamma$ Criteria}

Correlated 2.2 MeV $\gamma$ from neutron capture are distinguished
from accidental $\gamma$ from radioactivity by use of the likelihood
ratio, $R_\gamma$, which is defined to be the likelihood that the
$\gamma$ is correlated divided by the likelihood that the $\gamma$ is
accidental.  $R_\gamma$ depends on three quantities: the number of hit
PMTs associated with the $\gamma$ (the multiplicity is proportional to
the $\gamma$ energy), the distance between the reconstructed $\gamma$
position and positron position, and the time interval between the
$\gamma$ and positron (neutrons have a capture time in mineral oil of
186 $\mu$s, while the accidental $\gamma$ are uniform in
time). Fig. \ref{fig:Fig{10}_ruld} shows these distributions, which are
obtained from fits to the data, for both correlated 2.2 MeV $\gamma$
(solid curves) and accidental $\gamma$ (dashed curves). To determine
$R_\gamma$, the product of probabilities for the correlated
distributions is formed and divided by the product of probabilities
for the uncorrelated distributions. The accidental $\gamma$
efficiencies are measured from the laser-induced calibration events,
while the correlated $\gamma$ efficiencies are determined from the
Monte Carlo simulation of the experiment. Similar results for the
correlated $\gamma$ efficiencies are obtained from the cosmic-ray
neutron events, whose high energy gives them a slightly broader
position distribution.  The efficiencies for different $R_\gamma$
selections are shown in Table \ref{tab:rgamma}.  The systematic
uncertainty of these efficiencies is estimated to be $\pm 7\%$ of
their values.  Note that with the new reconstruction, the correlated
$\gamma$ efficiency has increased while the accidental $\gamma$
efficiency has decreased. For $R_\gamma >10$, the correlated and
accidental efficiencies are 0.39 and 0.003, respectively.  For the
previous reconstruction \cite{bigpaper2} the $R_\gamma^{old}>30$ cut
gave correlated and accidental efficiencies of 0.23 and 0.006,
respectively.

As checks of the likelihood distributions, Fig.  \ref{fig:Fig{11}_r2_nuec}
shows the $R_\gamma$ distributions for $\nu_e C \rightarrow e^-
N_{g.s.}$ exclusive events \cite{nuec}, where the $N_{g.s.}$ $\beta$
decays.  By definition, the $\nu_e C \rightarrow e^- N_{g.s.}$
reaction has no recoil neutron, so that its $R_\gamma$ distribution
should be consistent with a purely accidental $\gamma$ distribution. A
fit to the $R_\gamma$ distribution finds that the fraction of events
with a correlated $\gamma$, $f_c$, is $f_c = -0.004 \pm 0.007$
($\chi^2 = 4.6/9$ DOF).  Fig. \ref{fig:Fig{12}_r2_numu} shows the $R_\gamma$
distribution for the sample of $\mu^{\pm}$ events arising from the
reactions $\nu_\mu C \rightarrow \mu^- X$, $\bar \nu_\mu C \rightarrow
\mu^+ X$, and $\bar \nu_\mu p \rightarrow \mu^+ n$.  Correlated
$\gamma$ are expected for $\sim 14\%$ of these events \cite{kolbe}.  A
fit to the $R_\gamma$ distribution gives $f_c = 0.129 \pm 0.013$
($\chi^2 = 8.2/9$ DOF), in agreement with expectations.
Fig. \ref{fig:Fig{13}_numu_gamma} shows the distributions of $\Delta r$,
$\Delta t$, and $N_{hits}$ for events with $R_\gamma >1$ (left side)
and $R_\gamma <1$ (right side). The top plots show the distance
between the reconstructed $\gamma$ position and positron position,
$\Delta r$, the middle plots show the time interval between the
$\gamma$ and positron, $\Delta t$, and the bottom plots show the
number of hit PMTs associated with the $\gamma$, $N_{hits}$.
\subsection{Neutrino Oscillation Results}

Table \ref{tab:stats} shows the statistics for events that satisfy the
selection criteria for the primary $\bar \nu_\mu \rightarrow \bar
\nu_e$ oscillation search. An excess of events is observed over that
expected from beam-off and neutrino background that is consistent with
neutrino oscillations.  A $\chi^2$ fit to the $R_\gamma$ distribution,
as shown in Fig. \ref{fig:Fig{14}_r2_osc}, gives $f_c = 0.0567 \pm 0.0108$
($\chi^2 = 10.7/9$ DOF), which leads to a beam on-off excess of $117.9
\pm 22.4$ events with a correlated neutron.  Subtracting the neutrino
background from $\mu^-$ DAR followed by $\bar \nu_e p \rightarrow e^+
n$ scattering ($19.5 \pm 3.9$ events) and $\pi^-$ DIF followed by
$\bar \nu_\mu p \rightarrow \mu^+ n$ scattering ($10.5 \pm 4.6$
events) \cite{numubkgd} leads to a total excess of $87.9 \pm 22.4 \pm
6.0$ events, as shown in Table \ref{tab:table_le}.  This excess
corresponds to an oscillation probability of $(0.264 \pm 0.067 \pm
0.045)\%$, where the first error is statistical and the second error
is the systematic error arising from uncertainties in the backgrounds,
neutrino flux ($7\%$), $e^+$ efficiency ($7\%$), and $\gamma$
efficiency ($7\%$).  Note that our previously published result
\cite{bigpaper2}, based on the 1993-1995 data sample, was $(0.31 \pm
0.12 \pm 0.05)\%$.  Table \ref{tab:oscillation_probs} shows the effect
on the fitted oscillation probability of tightening some of the
selection criteria.

A clean sample of oscillation candidate events can be obtained by
requiring $R_\gamma >10$, where as shown in Table \ref{tab:stats}, the
beam on-off excess is $49.1 \pm 9.4$ events while the estimated
neutrino background is only $16.9 \pm 2.3$ events.
Fig. \ref{fig:Fig{15}_elec_gamma} shows the individual $\gamma$ distributions
for events with $20<E_e<60$ MeV and with $R_\gamma >1$ (left side) and
$R_\gamma <1$ (right side).  The top plots show the distance between
the reconstructed $\gamma$ position and positron position, $\Delta r$,
the middle plots show the time interval between the $\gamma$ and
positron, $\Delta t$, and the bottom plots show the number of hit PMTs
associated with the $\gamma$, $N_{hits}$.  
Fig. \ref{fig:Fig{16}_elec_en_rcut} displays the energy distribution of events with $R_\gamma >10$. The
shaded regions show the combination of neutrino background plus
neutrino oscillations at low $\Delta m^2$. The data agree well with
the oscillation hypothesis. As mentioned in section I, the 1993-1995
data runs employed a 30 cm water target, while the 1996-1998 data runs
used a high-Z metal target. A comparison of the energy distributions
of the two data samples is displayed in Fig. \ref{fig:Fig{17}_elec_en_rcut1},
which shows that the data samples are consistent within statistics.

Fig. \ref{fig:Fig{18}_elec_cos_rcut} shows the $\cos \theta_\nu$ distribution
for events with $R_\gamma >1$ and $36<E_e<60$ MeV. This energy range
is chosen because it is particularly clean with reduced neutrino
background, so that the $\bar \nu_e p \rightarrow e^+ n$ reaction
should dominate, while the $\gamma$ requirement is relaxed to increase
the statistics.  $\theta$ is the angle between the incident neutrino
and outgoing positron directions.  The shaded region in
Fig. \ref{fig:Fig{18}_elec_cos_rcut} shows the expected distribution from a
combination of neutrino background plus neutrino oscillations.  The
$<\cos \theta_\nu> = 0.04 \pm 0.12$, in agreement with the expectation
of $\sim 0.12$.

Figs. \ref{fig:Fig{19}_elec_space_rcut} ($D > 10$ cm ) and
\ref{fig:Fig{20}_elec_space_rcut2} ($D > 35$ cm ) show the spatial
distributions for events with $R_\gamma >10$ and $20<E_e<60$ MeV,
where z is along the axis of the tank (and approximately along the
beam direction), y is vertical, and x is transverse. The shaded
regions in Figs. \ref{fig:Fig{19}_elec_space_rcut} and
\ref{fig:Fig{20}_elec_space_rcut2} show the expected distributions from a
combination of neutrino background plus neutrino oscillations.
Fig. \ref{fig:Fig{21}_scat_xyz} shows
scatter plots of the x-y and y-z spatial distributions for events with
$R_\gamma >10$, $20<E_e<60$ MeV, and $D > 35$ cm.
Figs. \ref{fig:Fig{22}_elec_pid_rcut} and
\ref{fig:Fig{23}_elec_veto_rcut} show the
$\chi_{tot}^\prime$ and veto hit distributions for events with
$R_\gamma >10$ and $20<E_e<60$ MeV. The solid histogram in the veto
hit figure shows the distribution from $\nu_e C \rightarrow e^-
N_{g.s.}$ scattering.  
Finally, Fig. \ref{fig:Fig{24}_elec_loe_rcut} shows the
$L_\nu/E_\nu$ distribution for events with $R_\gamma >10$ and
$20<E_e<60$ MeV, where $L_\nu$ is the distance travelled by the
neutrino in meters and $E_\nu$ is the neutrino energy in MeV
determined from the measured positron energy and angle with respect to
the neutrino beam. The data agree well with the expectation from
neutrino background plus neutrino oscillations at low $\Delta m^2$
($\chi^2 = 4.9/8$ D.O.F.)  or high $\Delta m^2$ ($\chi^2 = 5.8/8$
D.O.F.).

\subsection{Tests of the $\bar \nu_\mu \rightarrow \bar \nu_e$ 
Oscillation Hypothesis}

A variety of tests of the $\bar \nu_\mu \rightarrow \bar \nu_e$
oscillation hypothesis have been performed.  One test of the
oscillation hypothesis is to check whether there is an excess of
events with more than one correlated $\gamma$. If the excess of events
is indeed due to the reaction $\bar \nu_e p \rightarrow e^+ n$, then
there should be no excess with more than one correlated $\gamma$
because the recoil $n$ is too low in energy ($<5$ MeV) to knock out
additional neutrons.  If, on the other hand, the excess involves
higher energy neutrons ($>20$ MeV) from cosmic rays or the beam, then
one would expect a large excess with $>1$ correlated $\gamma$, as
observed in the beam-off cosmic ray data.  However, as shown in Table
\ref{tab:gammas}, the excess of events with more than one correlated
$\gamma$ is approximately zero for both the full $20< E_e <60$ MeV
energy region and the lower background $36< E_e <60$ MeV energy
region, as expected for the reaction $\bar \nu_e p \rightarrow e^+ n$.

Another test of the oscillation hypothesis is to check the ``event lookback''
for events that satisfy the oscillation criteria in
order to ensure that the $\bar \nu_\mu p \rightarrow \mu^+ n$ background
is calculated correctly. The ``event lookback'' was installed prior to the 
1995 running and consisted of an extra trigger that read out all hit 
detector PMTs in the 6 $\mu$s interval before a primary event. Any
background just below the 18 hit muon threshold will clearly show up, 
especially in the hit range with $>11$ lookback hits, where the probability
of having an accidental lookback is only $5.6\%$.
However, as shown in Table \ref{tab:look}, 
the excess of primary events with $R_\gamma \ge 0$ or $R_\gamma > 10$
is consistent with the accidental lookback probability. 
Thus, this ``event lookback'' check provides
additional assurance that the $\bar \nu_\mu p \rightarrow \mu^+ n$ background
calculation of 10.5 events is not underestimated.

Both major backgrounds with a correlated neutron arise initially from
$\pi^-$ DIF. Therefore, a final test of the oscillation hypothesis is
to check whether the $\bar \nu_\mu$ flux from $\pi^-$ DIF is
correct. However, this has already been tested by the fit to the
$R_\gamma$ distribution, discussed above, of $\nu_\mu C \rightarrow
\mu^- X$, $\bar \nu_\mu C \rightarrow \mu^+ X$, and $\bar \nu_\mu p
\rightarrow \mu^+ n$ inclusive events \cite{numuc}.  For these
reactions, correlated $\gamma$ are expected for $\sim 14\%$ of the
events \cite{kolbe}, due mainly to the $\bar \nu_\mu$ flux.  A fit to
the $R_\gamma$ distribution gives $f_c = 0.129 \pm 0.013$ ($\chi^2 =
8.2/9$ DOF), in agreement with
expectation. Fig. \ref{fig:Fig{25}_elec_pdtmin2} shows the time to the
previous event for $R_\gamma>10$ electron events prior to applying the
$\Delta t_{past}>12\mu$s selection. The top graph in the figure shows
that the beam excess events are in agreement with our expectations for
the $\nmbp$ and $\nmbc$ channels. With the same data on a smaller vertical
scale, the bottom graph shows events with accidental past activities,
in agreement with expectations from random cosmic ray backgrounds.

%
%
%
\section{The Decay-in-Flight Oscillation Analysis}

The high energy $\nm$s from $\pi^+$ decay-in-flight are a potential
source of $\nmne$ oscillation events.  The contamination of $\nue$s
from standard sources is small, at the level of $0.1\%$. Unfortunately,
the cosmic-ray backgrounds are large, with the result that the
fluctuations in the beam-on, beam-off subtraction are comparable to
the expected signal.  Prior to 1996, it was realized that the
1996-1998 data would not support a stand alone, decay-in-flight
analysis due to the larger beam-off backgrounds that are inherent in
running with a heavy target.  However, the analysis presented here is
extended up to an electron energy of 200 MeV because the
decay-in-flight data constrain the region $>2$ eV$^2$, especially
around $6 $eV$^2$.

The above analysis is applied to data in the energy range $60<E_e<200$
MeV , with the additional requirement that there be no associated
$\gamma$.  This sample is only sensitive to $\nu_\mu \rightarrow \nu_e$
oscillations, and results in a beam on-off excess of $14.7 \pm
12.2$ events. The signal expected for $100\%$ $\nu_\mu$ to $\nu_e$
transmutation is estimated to be 7800 events, and the $\nu_e$
background from $\mu^+ \rightarrow e^+ \bar \nu_\mu \nu_e$, $\pi^+
\rightarrow e^+ \nu_e$, and $\nu e \rightarrow \nu e$ is estimated to
be $6.6 \pm 1.7$ events, resulting in a total excess of $8.1 \pm 12.2
\pm 1.7$ events or an oscillation probability of $(0.10 \pm 0.16 \pm
0.04)\%$, as shown in Table \ref{tab:table_he}. This result is lower
than but consistent with our higher precision analysis of the
1993-1995 data sample \cite{paper3}.  That analysis determined the
selection criteria by maximizing the acceptance divided by the square
root of the beam-off background, which produced much less beam-off
background overall. It gave a total excess of $18.1 \pm 6.6 \pm 4.0$
oscillation events, corresponding to an oscillation probability of
$(0.26 \pm 0.10 \pm 0.05)\%$.  Due to changes in the neutrino
production target, the 1996-1998 data sample had reduced DIF flux and
higher beam-off background compared to the 1993-1995 data. Based on
our DAR oscillation result and assuming that CP is conserved in the
lepton sector, we would expect the DIF oscillation probability to be
$\sim 0.26\%$ at high $\Delta m^2$ (where $(1.27 \Delta m^2 L_\nu
/E_\nu) >> 1$) and $\sim 0.05\%$ at low $\Delta m^2$ (where $(1.27
\Delta m^2 L_\nu /E_\nu) << 1$).

The $0.16\%$ statistical error on the oscillation probability in the
present analysis is larger than the $0.10\%$ statistical error of the
previous analysis. That is because the present analysis uses the
electron selection criterion developed for the DAR region
below 60 MeV, which is less effective in removing the background to
electron events in the DIF region above 60 MeV.  When
compared to the previously published DIF analysis, the
beam-off background for these higher energy events is 3.42 times
higher in the present analysis, while the number of expected events
for $100\%$ $\nmne$ transmutation is only 1.16 times higher. The
previous analysis observed a 2.6 sigma excess, compared to the 0.6
sigma excess of the present analysis.

%
%
%
\section{The Neutrino Oscillation Fit}
\subsection{Introduction}

We describe in this section the $(\sin^22\theta,\Delta m^2)$
likelihood ($\mathcal{L}$) fitter. The fitter is applied to beam-on
events in the final oscillation sample and calculates a likelihood in
the $(\sin^22\theta,\Delta m^2)$ plane in order to extract the favored
oscillation parameters.  The fit is similar to that performed in
reference \cite{lanl-eprint-ke}.

The $\mathcal{L}$ product in the $(\sin^22\theta,\Delta m^2)$ plane is
formed over the individual beam-on events that pass the oscillation
cuts.  This three-dimensional contour is sliced to arrive finally at
the LSND allowed oscillation region.  The beam-related backgrounds are
determined from MC event samples for each individual background
contribution. The MC contains the trigger simulation and generally
very well reproduces the tank response to all particles of
interest. Agreement between the data and MC is excellent.  The fit is
over the entire electron energy range $20<E_e<200$ MeV. Therefore, DIF
oscillations and DIF backgrounds in addition to the usual DAR
processes are considered.

\subsection{Formalism}

Each beam-on event is characterized by four variables: the electron
energy, $E_e$, the electron reconstructed distance along the tank
axis, $z$, the reconstructed direction cosine the electron makes with
the neutrino, $\cos\theta_\nu$, and the likelihood ratio that the
event has a correlated 2.2 MeV $\gamma$, $R_\gamma$.  Each of the
neutrino-induced background processes is simulated, and the simulation
is compared to real events in the detector.  Accidental $\gamma$
events are used with real neutrino processes to simulate accidental
events. Beam-off events are used as a background contribution after
scaling by the measured time-dependent duty factor.  The duty factor
for this analysis was determined by using the entire raw event sample
to measure the ratio of beam-on time to beam-off time.  The raw event
sample consists mostly of beta-decay events and is, to a good
approximation, unbiased by beam-related events.  The duty factor for
each run was determined by dividing the number of raw events when the
beam-on bit was set by the number of raw events where it was not set.
This resulted in a duty factor for each run that was used to weight
the beam-off events to determine the beam-unrelated subtraction for
the final event sample.

For every point in the $(\sin^22\theta,\Delta m^2)$ plane, oscillation
signal events are generated to complete the description of sources
expected in the beam-on sample. There are 5697 beam-on events in the
data sample, and a likelihood is calculated for each one based on the
values of $E_e$, $z$, $\cos\theta_\nu$ and $R_\gamma$.

Formally, each neutrino beam-on event $j$ is assigned a probability
$p_j(E_e,z,\cos\theta_\nu,R_\gamma)$ equal to a sum of probabilities
$q_i(E_e,z,\cos\theta_\nu,R_\gamma)$ from the backgrounds plus
oscillations. It then remains to add the $q_i$ with expected
fractional contributions $r_i$ and take the product over all the
beam-on events. The likelihood is thus
\begin{equation}
\mathcal{L} =  (\prod_{j=1}^{N_{beam-on}} p_j),
\end{equation}
where
\begin{equation}
p_j(E_{ej},R_{\gamma j},\cos\theta_{\nu j},z_j) = \sum_{i=1}^{N_{contributions}} q{_i}(E_{ej},R_{\gamma j},\cos\theta_{\nu j},z_j)
\cdot r_i.
\end{equation}
Additionally, two normalization requirements must hold:
\begin{equation}
\sum_{i=1}^{N_{contributions}} r_i = 1,
\end{equation}
and
\begin{equation}
\int dE_e\,dR_\gamma\,d(\cos\theta_\nu)\,dz\,\, q_i(E_e,R_\gamma,\cos\theta_\nu,z) = 1
\end{equation}
for each contribution, $i$.  Together, these requirements ensure that
every observed beam-on event has a probability of occurrence equal to
1.

\subsection{Background Variation}

It is necessary to allow for the fact that the backgrounds are not
perfectly known.  The background variation is performed by calculating
the above likelihood at each point in the $(\sin^2 2\theta,\Delta
m^2)$ plane many times, varying over the expected $\sigma$ for each
background.  For each background configuration, the $\mathcal{L}$ is
weighted with a Gaussian factor for each background that is off its
central value.  The background configurations are varied so that the
beam-unrelated background (BUB) varies independently and the
beam-related backgrounds (BRBs) are locked together. Different
background varying procedures give very similar results.

\subsection{The expression for the Likelihood}

Finally, the likelihood can be expressed as
\begin{equation}
\mathcal{L} = \int\mathcal{D}N_{bgd}
\exp(-(N_{bgd}-N_{bgd,exp})^2/2\sigma^2) \cdot  (\prod_{i=1}^{N_{beam-on}} p_i), 
\end{equation}
where the $\int\mathcal{D}N_{bgd}$ represents, schematically, the
background variation described above.

	\subsection{The Input}

The $q_i$ for each of the background and signal processes are all
generated from the MC, except for the BUB $q_i$, which is generated
from the beam-off data events.  There are separate MC runs for each of
the above BRB processes.  Some of these backgrounds are grouped
together (appropriately weighted) into a few common $q_i$s for easier
bookkeeping, as indicated in Table \ref{tab:tb2}. This is done for
backgrounds which don't need to be separately varied.  Several small,
beam-related backgrounds, DIF $\nu e \rightarrow \nu e$ elastic
scattering and $\pi^+\rightarrow e^+ \nu_e$ DAR followed by $\nu_e C
\rightarrow e^- N$ scattering, are contained in their DAR and DIF
counterparts.

	\subsection{Slicing the contour}
		\label{slices}

		\subsubsection{The Feldman-Cousins Method}

The Feldman-Cousins method 
\cite{feldman-cousins} can be applied to the LSND $\cal{L}$
contour in the following way.  At a particular point in the
$(\sin^22\theta,\Delta m^2)$ plane, create thousands of generated data
sets comprised of background and oscillations.  For each Monte Carlo
experiment compute $\delta L = L_{Max} - L_{MC}$, where $L =
log\cal{L}$, $L_{MC}$ is $L$ at the particular point in the
$(\sin^22\theta,\Delta m^2)$ plane assumed in the Monte Carlo, and
$L_{Max}$ is the log likelihood at the values of $\sin^22\theta$ and
$\Delta m^2$ that maximize $\cal{L}$.  From a histogram of $\delta L$
for the thousands of Monte Carlo data sets one obtains the selection
that contains, for example, $90\%$ of the experiments.  Finally,
determine this selection at many points in the $(\sin^22\theta,\Delta
m^2)$ plane. The resulting function of $\sin^22\theta$ and $\Delta
m^2$ corresponds to the $90\%$ C.L. allowed LSND region.

This approach, as practiced in reference \cite{lanl-eprint-ke},
required large amounts of CPU.  Even scanning a judiciously chosen
$(\sin^22\theta,\Delta m^2)$ region is CPU intensive, and setting up
and running the generated data sets would take many months.
Therefore, the full Feldman-Cousins method will not be followed here.
As shown below, using slices derived from a different LSND data set to
determine the $\mathcal{L}$ contours for this data set, the results
obtained with the Feldman-Cousins method are similar to other methods.

		\subsubsection{The Bayes Method}

For the Bayes method one presumes a prior expectation of the
oscillation parameters from 0.01 to 100.0 eV$^2$ in $\Delta m^2$ and
0.001 to 1.0 in $\sin^22\theta$. The assumption of this prior
expectation is what makes this approach Bayesian.  Each bin in the
$(\sin^22\theta,\Delta m^2)$ plane is assigned a weight $w$, where $w
= \delta x \, \delta y \cdot \mathcal{L}$.  That is, the weight is the
measure of the probability distribution times the $\mathcal{L}$. The
measure $\delta x\, \delta y$ is taken to be $\delta
(\ln\sin^22\theta) \delta (\ln\Delta m^2) $. The $90\%$ and $99\%$
C.L. regions are then determined by integrating over the
$(\sin^22\theta,\Delta m^2)$ plane.

		\subsubsection{The Constant-Slice Method}

The constant-slice method makes a slice at a constant value of $L$.
If, for example, the log likelihood were a two-dimensional Gaussian,
slices of 2.3 and 4.6 units down from the peak $L$ would correspond to
$90\%$ and $99\%$ C.L., respectively.  Fig. \ref{fig:Fig{26}_cl90} shows that the
Feldman-Cousins, Bayesian, and constant-slice methods all give about
the same $90\%$ regions.  Note that for the Feldman-Cousins method the
slices are derived from a different LSND data set.  We use the
constant-slice method in this paper to denote the favored regions in
the $(\sin^22\theta,\Delta m^2)$ plane.

\subsection{Statistical Issues and Technical Hurdles}
		\label{stats}

Preserving correlations in the $E_e,R_\gamma,\cos\theta_\nu,z$
parameter space over which the $\mathcal{L}$ fit is performed is
sometimes difficult, due to the fact that for certain backgrounds the
3600 bin parameter space is too large to characterize. In particular,
$R_\gamma$, with its logarithmic behavior for backgrounds in which
uncorrelated $\gamma$s are present, is especially difficult.  This
problem was resolved for the BUB by binning the other parameters very
coarsely, effectively ignoring correlations in some regions of the
four-dimensional parameter space.  Statistical problems with the MC
BRB sample, in which uncorrelated $\gamma$s are present, were dealt
with in a similar manner. Such measures were safe approximations for
the fiducial volume of interest.

Other technical difficulties in certain ranges of $\Delta m^2$ were
overcome with weighting techniques. The origin of the difficulties was
always one of limited statistical samples that characterized the
probability distribution functions for the backgrounds. Another
problem involved re-weighting by $\sin^2 (1.27 \Delta m^2 {L_\nu \over
E_\nu})$, which required prohibitive numbers of MC events and the
simultaneous breaking of correlations in the four-dimensional space.
However, these difficulties were overcome by smearing $L_\nu$, the
distance travelled by the neutrino, and $E_\nu$, the neutrino energy,
with the Gaussian widths determined from the position and energy
resolutions.

	\subsection{Results}

A $(\sin^22\theta,\Delta m^2)$ oscillation parameter fit for the
entire data sample, $20<E_e<200$ MeV, is shown in
Fig. \ref{fig:Fig{27}_lhd}. The fit includes both $\bar \nu_\mu \rightarrow
\bar \nu_e$ and $\nu_\mu \rightarrow \nu_e$ oscillations, as well as
all known neutrino backgrounds. The inner and outer regions correspond
to $90\%$ and $99\%$ CL allowed regions, while the curves are $90\%$ CL
limits from the Bugey reactor experiment \cite{bugey}, the CCFR
experiment at Fermilab \cite{ccfr}, the NOMAD experiment at CERN
\cite{nomad}, and the KARMEN experiment at ISIS \cite{karmen2}.  The
most favored allowed region is the band from $0.2 - 2.0$ eV$^2$,
although a region around $7$ eV$^2$ is also possible, but has been
made less probable by the $\nu_\mu \rightarrow \nu_e$ analysis.

The KARMEN experiment also searches for $\bar \nu_\mu \rightarrow \bar
\nu_e$ oscillations with a detector that is similar to LSND. A
comparison of the two experiments is given in Table \ref{tab:LSNDKAR}.
LSND is a more massive detector, has a higher intensity neutrino
source, and has good particle identification, while KARMEN has better
energy resolution and the advantage of a much lower duty factor that
helps eliminate cosmic-ray events.  In addition, KARMEN is located
17.5 m from the neutrino source, compared with 30 m for
LSND. Therefore, the experiments have sensitivities that peak at
different values of $\Delta m^2$.  At low $\Delta m^2$, for example,
an experiment at 30 m is 2.94 times more sensitive to neutrino
oscillations than an experiment at 17.5 m. Note that a global analysis
of the two experiments was performed by Eitel \cite{lanl-eprint-ke}
using intermediate data sets.

The event breakdown from the $20<E<200$ MeV four-dimensional fit is
shown in Table \ref{tab:tb2} at the best-fit point
\[
(\sin^22\theta, \Delta m^2)_{best-fit} = (0.003, 1.2 \rm{eV}^2).
\] 
The number of $\bar \nu_\mu \rightarrow \bar \nu_e$ oscillation events
at the best-fit point is 89.5 events, which agrees well with the $87.9
\pm 22.4 \pm 6.0$ event excess from the fit to the $R_\gamma$
distribution.  The whole low $\Delta m^2$ region gives an almost
equally good fit within 0.5 log-likelihood units.  Projections onto
$E_e,R_\gamma,z,\cos\theta_\nu$ from the four-dimensional fit at the
best fit value of $(\sin^22\theta,\Delta m^2)$ are plotted in
Fig. \ref{fig:Fig{28}_projections}.  The fit is relatively insensitive to the
starting values and gives good overall agreement wth the data.

%
%
%
\section{Conclusions}

The final LSND $\bar \nu_\mu \rightarrow \bar \nu_e$ oscillation
results are presented for all six years of data collection,
1993-1998. The analysis employed a new event reconstruction that
greatly improved the correlation of the $e^+$ and 2.2 MeV $\gamma$
from the reaction $\bar \nu_e p \rightarrow e^+ n$, thus greatly
reducing the background from neutrino events followed by an accidental
$\gamma$. These final results are consistent with our earlier
analysis of the 1993-1995 data sample \cite{bigpaper2}; in particular,
the results from the 1993-1995 data sample, which used a water target,
are consistent with the results from 1996-1998, which made use of a
high-Z target.  

A global fit was performed to all event categories shown in Table
\ref{tab:GFcategories} in order to check our understanding of neutrino
processes in the experiment.  The parameters resulting from this fit,
shown in Table \ref{tab:GFparameters}, together with neutrino
oscillations, yield a good description of all the observed data.

The LSND experiment provides evidence for neutrino oscillations from
the primary $\bar \nu_\mu \rightarrow \bar \nu_e$ oscillation search.
A total excess of $87.9 \pm 22.4 \pm 6.0$ $\bar\nu_e p \rightarrow
e^+n$ events with $e^+$ energy between 20 and 60 MeV is observed above
expected neutrino-induced backgrounds.  This excess corresponds to an
oscillation probability of $(0.264 \pm 0.067 \pm 0.045)\%$.  A fit to
all of the LSND neutrino processes determines the allowed oscillation
parameters in a two--generation model.  In conjunction with other
available neutrino oscillation limits, the LSND data suggest that
neutrino flavor oscillations occur with a $\Delta m^2$ in the range
$0.2-10$ eV$^2$/c$^4$.  

In addition, using the same event selection,
results are also presented for the decay-in-flight 
energy region. Although a clear event excess is not observed, the
results are consistent with the $\bar \nu_\mu \rightarrow
\bar \nu_e$ oscillation signal and with our higher precision analysis of
the 1993-1995 data sample \cite{paper3}, which determined the
selection parameters by maximizing the acceptance divided by the
square root of the beam-off background and which had much less
beam-off background overall.

At present, the LSND results remains the only evidence for
appearance neutrino oscillations and implies that at least one
neutrino has a mass greater than $0.4$ eV/c$^2$.  The MiniBooNE
experiment at Fermilab \cite{loi}, which is presently under
construction, is expected to provide a definitive test of the LSND
results, and if the neutrino oscillation results are confirmed, will
make a precision measurement of the oscillation parameters.

\paragraph*{Acknowledgments}

This work was conducted under the auspices of the US Department of Energy,
supported in part by funds provided by the University of California for
the conduct of discretionary research by Los Alamos National Laboratory.
This work is also supported by the National Science Foundation.
We are particularly grateful for the extra effort that was made by these
organizations to provide funds for running the accelerator at the end of
the data taking period in 1995.
It is pleasing that a large number of undergraduate students
from participating institutions were able to contribute 
significantly to the experiment. We thank K. Eitel for his valuable
contributions to the analysis.

%
%
\clearpage

\newcommand{\taboptions}{tbp}
%
%
\begin{table*}[\taboptions]
\vspace*{\fill}
\caption{
The proton beam statistics for each of the 
years of running, 1993 through 1998.
}
\vspace{10mm}
\begin{tabular} {|c|c|c|c|c|}
\hline
Year &Charge (C)&Protons ($\times 10^{22}$) & A6 target & Active Targets  \\ \hline
1993 &     1787	   &    1.12     &   water     &   A1, A2, A6  \\
1994 & 	   5904	   &    3.69     &   water     &   A1, A2, A6  \\
1995 & 	   7081	   &    4.42 	 &  water &  A1, A2, A6  \\
1996 & 	   3789	   &   2.37      & high-Z metal &   A6 \& partial A2\\
1997 & 	   7181	   &   4.48   	 & high-Z metal &   A6 only   \\
1998 & 	   3154	   &   1.97	 & high-Z metal &   A6 only   \\
\hline
\end{tabular}
\label{tab:proton_beam}
\end{table*}

%
\begin{table*}[\taboptions]
\vspace*{\fill}
\caption{
The $\mu^-$ absorption rates for materials in the target area
\cite{bib:muabsorp1}.
}
\vspace{10mm}
\begin{tabular} {|c|c|c|}
\hline
Material&Z     & $\mu^-$ Absorption Rate ($\mu s^{-1}$) \\ 
\hline
  H   &1	&$0.00042 \pm 0.00002$ \\
  Be  &4       &  $0.0074 \pm 0.0005$ \\
  C   &6      &   $0.0388 \pm 0.0005$ \\
  O   &8     &   $0.1026 \pm 0.0006$ \\
  Al  &13       &  $0.7054 \pm 0.0013$ \\
  Fe  &26       &  $4.411 \pm 0.024$ \\
  Cu  &29         &   $5.676 \pm 0.037$ \\
  Zn  &30        & $5.834 \pm 0.039$ \\
  Mo  &42       &  $9.61 \pm 0.15$ \\
  Ta  &73       &  $12.86 \pm 0.13$ \\
  Pb  &82       &   $13.45 \pm 0.18$ \\
  U   &92       &  $12.60 \pm 0.04$ \\
\hline
\end{tabular}
\label{tab:muabsorp}
\end{table*}


\begin{table*}[\taboptions]
\caption{Average neutrino fluxes in LSND. Both decay at rest (DAR)
and decay in flight (DIF) are shown in $\nu/\rm{cm}^2$. The $\nu_\mu$
and $\bar \nu_\mu$ DIF fluxes are above $\mu$ production threshold.}
\newcommand{\m}{\hphantom{$-$}}
\newcommand{\cc}[1]{\multicolumn{1}{c}{#1}}
\renewcommand{\tabcolsep}{2pc} 
\renewcommand{\arraystretch}{1.2} 
\vspace{10mm}
\begin{tabular}{|c|c|c|c|c|}
\hline
Source      & Type              & 1993-1995 Flux      & 1996-1998 Flux      & Total Flux \\ \hline

$\mu^+$ DAR & $\nmb$ and $\nue$ & $\ee{7.38}{13}$ & $\ee{5.18}{13}$ & $\ee{1.26}{14}$ \\
$\mu^-$ DAR & $\nm$ and $\neb$  & $\ee{5.96}{10}$ & $\ee{4.87}{10}$ & $\ee{1.08}{11}$ \\
$\pi^+$ DIF & $\nm$             & $\ee{1.37}{12}$ & $\ee{8.26}{11}$ & $\ee{2.20}{12}$ \\
$\pi^-$ DIF & $\nmb$            & $\ee{1.45}{11}$ & $\ee{1.11}{11}$ & $\ee{2.56}{11}$ \\
$\pi^+$ DIF & $\nue$            & $\ee{5.56}{ 8}$ & $\ee{5.01}{ 8}$ & $\ee{1.06}{ 9}$ \\
$\mu^+$ DIF & $\nue$            & $\ee{4.13}{ 9}$ & $\ee{2.44}{ 9}$ & $\ee{6.57}{ 9}$ \\
\hline
\end{tabular}\\[2pt]
\label{tab:fluxes}
\end{table*}
%
%
%
\begin{table*}[\taboptions]
\vspace*{\fill}
\caption{
Cross section uncertainties for the neutrino reactions 
with two-body final 
states that occur in LSND.
The cross sections for these processes are known accurately
because either related measurements can be used to constrain
the matrix elements or only fundamental
particles are observed. Also shown are the corresponding neutrino
flux constraints.
}
\vspace{10mm}
\begin{tabular} {|c|c|c|c|c|}
\hline
Process&$\sigma$ Constraint  & $\sigma$ Uncertainty &Flux Constraint \\
\hline
$\nu e \rightarrow \nu e$ & Standard Model Process  &  $1\%$&
$\mu^+\rightarrow\nu_e\bar\nu_\mu e^+$ DAR \\
$^{12}C(\nu_e,e^-)^{12}N_{g.s.}$     & $^{12}N_{g.s.}$ &  $5\%$&
$\mu^+\rightarrow\nu_e\bar\nu_\mu e^+$ DAR \\
$^{12}C(\nu_\mu,\mu^-)^{12}N_{g.s.}$ & $^{12}N_{g.s.}$ &  $5\%$&
$\pi^+\rightarrow\nu_\mu \mu^+$ DIF \\
$p(\bar\nu_\mu,\mu^+)n$& neutron decay&  $5\%$&
$\pi^-\rightarrow\bar\nu_\mu \mu^-$ DIF \\
\hline
\end{tabular}
\label{tab:2B-reactions}
\end{table*}

\begin{table*}[\taboptions]
\caption{
The average efficiencies for electrons in the fiducial volume with
energies in the range $20 <E_e<60$ MeV.}
\vspace{10mm}
\begin{tabular}{|c|c|} 
\hline 
Criteria & Efficiency \\ 
\hline
\multicolumn{2}{|c|}{Electron Reduction}\\
\hline 
Energy $> 15$ MeV& $1.00$ \\
Veto Hits $< 4$&$0.98 \pm 0.01$ \\
No Laser Tag& $1.00$ \\
Loose Electron PID&$0.96 \pm 0.01$ \\
Vertex $> 10$ cm from PMTs&$1.00$ \\
Cosmic Muon Cut &$0.92 \pm 0.01$ \\
\hline
\multicolumn{2}{|c|}{Electron Selection}\\
\hline
$\Delta t_{past}>12\mu$s        &  $0.96 \pm 0.01$  \\
$\Delta t_{future}>8\mu$s       &  $0.99 \pm 0.01$  \\
No bottom veto hit               & $1.00$ \\
$-1.5<\chi_{tot}^\prime<0.5$            &  $0.84 \pm 0.01$ \\
$0.3<\chi^{old}_{tot}<0.65$  (1993 only)      &  $0.98 \pm 0.01$    \\
$85\rm{ns}<t_{event}<210\rm{ns}$           &  $1.00$  \\
$\Delta t^{best}_{veto}>30\rm{ns}$    &  $0.97 \pm 0.01$   \\
D $>35$ cm & $0.88 \pm 0.02$      \\
$N_\gamma <1$, $E>60$ & $1.00$    \\
$N_\gamma <2$, $E<60$ & $1.00$ \\
\hline
\multicolumn{2}{|c|}{Deadtime}\\
\hline
DAQ \& Tape Deadtime & $0.96  \pm 0.02$ \\
Veto Deadtime & $0.76 \pm 0.02$ \\
\hline
Total&$0.42 \pm 0.03$   \\
\hline
\end{tabular}
\label{tab:electron_selection}
\end{table*}

%
\begin{table*}[\taboptions]
\vspace*{\fill}
\caption{
Event categories used to determine the number of events from 
standard neutrino processes.
}
\vspace{10mm}
\begin{tabular} {|c|c|c|c|}
\hline
Category         & Past Event & Primary Event   &   Future Event \\
\hline
$e$     &  -         & $\nu_e$         &       -          \\
$e\ \beta$       &  -         & $\nu_e$         & $^{12}N$ decay   \\
$e\ \gamma$      &  -         & $\nu_e$         & $n$ capture      \\
$\mu\ e$         &  $\mu$ & $e$(muon decay) &       -          \\
$\mu\ e\ \beta$  &  $\mu$ & $e$(muon decay) & $^{12}N$ decay   \\
$e\ \gamma\ \beta$&  - & $\nu_e$ &accidental $\gamma$ + $^{12}N$ decay \\
$\mu\ e\ \gamma$ &  $\mu$ & $e$(muon decay) &  $n$ capture     \\
$e$ no $\beta$   &  -         & $\nu_e$         &     -           \\ 
\hline
\end{tabular}
\label{tab:GFcategories}
\end{table*}

\begin{table*}[\taboptions]
\vspace*{\fill}
\caption{
Parameters adjusted during the least squares fit procedure, 
along with the fitted correction values, central correction values,
and nominal parameter values.
}
\vspace{10mm}
\small
\begin{tabular} {|c|c|c|c|}
\hline
Parameter       & Fitted Correction Value& Central Correction Value&
Nominal Parameter Value \\
\hline
\multicolumn{4}{|c|}{Flux Parameters}\\
\hline

  $\Phi_{DIF}$ & $  0.88 \pm   0.09 $ & 1.00 $\pm$ 0.15 & $0.22\times 10^{13}
\nu/cm^2$ \\
  $\Phi_{DAR}$ & $  1.01 \pm   0.05 $ & 1.00 $\pm$ 0.07 & $12.6\times 10^{13}
\nu/cm^2$\\
  ${\pi^-\over\pi^+}\ ratio$ & $0.90\pm   0.19 $ & 1.00 $\pm$ 0.10 & 0.12 \\
\hline
\multicolumn{4}{|c|}{Cross Section Parameters}\\
\hline
  $\sigma(\nu_\mu ^{12}C\rightarrow \mu^-\ ^{12}N^*)$ & $0.68 \pm 0.23$ & 1.00 $\pm$ 0.25 & $15.2\times 10^{-40}\ cm^2$\\
  $\sigma(\bar\nu_\mu p \rightarrow \mu^+ n)$ & $ 0.97 \pm   0.05$ & 1.00 $\pm$ 0.05 & $4.9\times 10^{-40}\ cm^2$\\
  $\sigma(\nu_e ^{12}C\rightarrow e^-\ ^{12}N_{g.s.})$ & $ 1.01 \pm   0.05$
& 1.00 $\pm$ 0.05 &$9.2\times 10^{-42}\ cm^2$ \\
  $\sigma(\nu_e ^{12}C\rightarrow e^-\ ^{12}N^*)$ & $1.02 \pm 0.13  $ & 1.00 $\pm$ 0.25 & $4.1\times 10^{-42}\ cm^2$\\
  $\sigma(\nu_e ^{13}C \rightarrow e^-\ ^{13}N)$ & $ 0.93 \pm 0.28$ & 1.00 $\pm$ 0.30 & $0.53\times 10^{-40}\ cm^2$\\
\hline
\multicolumn{4}{|c|}{Efficiency Parameters}\\
\hline
  $\epsilon_\mu$ & $   1.00 \pm   0.06 $ & $1.00 \pm 0.07$ & $0.93$ \\
  $\epsilon_\beta$ & $   1.00 \pm   0.04$ & $1.00 \pm 0.07$ & $0.65$ \\
  $\epsilon_e$ & $   1.00 \pm   0.05 $ & $1.00 \pm 0.07$ & $0.42$ \\
  $\epsilon_\gamma$ & $   0.91 \pm   0.03 $ & $1.00 \pm 0.07$ & $0.60$ \\
  duty ratio & $   0.95 \pm   0.03$ & $1.00 \pm 0.03$ & $0.060$ \\
\hline
\end{tabular}
\normalsize
\label{tab:GFparameters}
\end{table*}

\begin{table*}[\taboptions]

\caption{The estimated number of events in the
$20<E_e<60$ MeV energy range due to $100\%$
$\bar \nu_\mu \rightarrow \bar \nu_e$ transmutation
and to the two beam-related backgrounds with neutrons,
$\mu^-$ decay at rest in the beam stop followed
by $\bar \nu_e p \rightarrow e^+ n$ scattering in the detector
and $\pi^-$ decay in flight in the beam stop followed by
$\bar \nu_\mu p \rightarrow \mu^+ n$ scattering.
The $\pi^-$ DIF background includes contributions
from $\bar \nu_\mu C \rightarrow \mu^+ n X$
and $\nu_\mu C \rightarrow \mu^- n X$ scattering, as well
as a small $\bar \nu_e$ background from $\pi^-$ and $\mu^-$ DIF.
The events must satisfy the electron selection criteria, but
no correlated $\gamma$ requirement is imposed.}

\vspace{10mm}

\begin{tabular}{|c|c|c|}
\hline
Neutrino Source&Reaction&Number of Events \\
\hline
$\mu^+$ DAR&$100\%$ $\bar \nu_\mu \rightarrow \bar \nu_e$&$33300 \pm 3300$ \\
$\mu^-$ DAR&$\bar \nu_e p \rightarrow e^+ n$&$19.5 \pm 3.9$ \\
$\pi^-$ DIF&$\bar \nu_\mu p \rightarrow \mu^+ n$&$10.5 \pm 4.6$ \\
\hline
\end{tabular}
\label{tab:sigback}
\end{table*}

\begin{table*}[\taboptions]

\caption{
The correlated and accidental $\gamma$
efficiencies for different $R_\gamma$
selections. The systematic
uncertainty of these efficiencies is estimated to be $\pm 7\%$
of their values.
}

\vspace{10mm}

\begin{tabular}{|c|c|c|}
\hline
Selection &Correlated $\gamma$ Efficiency&Accidental $\gamma$ Efficiency \\
\hline
$R_\gamma>1$&0.51&0.012 \\
$R_\gamma>10$&0.39&0.0026 \\
$R_\gamma>100$&0.17&0.0002 \\
\hline
\end{tabular}
\label{tab:rgamma}
\end{table*}

\clearpage

\begin{table*}[\taboptions]

\caption{Numbers of beam-on events that satisfy the selection
criteria for the primary $\bar \nu_\mu \rightarrow \bar \nu_e$
oscillation search with $R_\gamma>1$, $R_\gamma>10$, and
$R_\gamma>100$. Also shown are
the beam-off background, the
estimated neutrino background, the excess of events that
is consistent
with neutrino oscillations, and the probability that the excess
is due to a statistical fluctuation.}

\vspace{10mm}

\begin{tabular}{|c|c|c|c|c|c|}
\hline
Selection &Beam-On Events&Beam-Off Background&$\nu$ Background&
Event Excess&Probability \\
\hline
$R_\gamma>1$ &205&$106.8\pm2.5$&$39.2 \pm 3.1$&$59.0\pm14.5 \pm 3.1$
&$7.8\times 10^{-6}$ \\
$R_\gamma>10$ &86&$36.9\pm1.5$&$16.9 \pm 2.3$&$32.2\pm9.4 \pm 2.3$
&$1.1 \times 10^{-4}$ \\
$R_\gamma>100$ &27&$8.3\pm0.7$&$5.4 \pm 1.0$&$13.3\pm5.2 \pm 1.0$
&$1.8 \times 10^{-3}$ \\
\hline
\end{tabular}
\label{tab:stats}
\end{table*}

\begin{table*}[\taboptions]
\caption{The number of excess events in the $20<E_e<60$ MeV
energy range, together with the corresponding oscillation
probability if the excess is due to
$\bar \nu_\mu \rightarrow \bar \nu_e$ oscillations.
Also shown are the results from the analysis of
the 1993-1995 data sample \cite{bigpaper2}.
}
\vspace{10mm}
\begin{tabular}{|c|c|c|}
\hline
Analysis& Excess Events&Oscillation Probability \\
\hline
Present Analysis (1993-1998)&$87.9 \pm 22.4 \pm 6.0$
&$(0.264\pm 0.067 \pm 0.045)\%$ \\
Previous Analysis (1993-1995)&$51.0 ^{+20.2}_{-19.5} \pm 8.0$
&$(0.31 \pm 0.12 \pm 0.05)\%$ \\
\hline
\end{tabular}
\label{tab:table_le}
\end{table*}

\begin{table}[ht]
\caption{
The oscillation probabilities obtained with various selections. The 
$S>0.5$ selection was used in the previous analysis \cite{bigpaper2}.
The nominal values are shown in Table \ref{tab:table_le}}.
\vspace{10mm}
\begin{tabular}{|c|c|} 
\hline 
Selection & Oscillation Probability \\ 
\hline
Nominal&$(0.264 \pm 0.067 \pm 0.045)\%$ \\
Nominal + $\Delta t_{past}>20\mu$s  & $(0.220 \pm 0.064 \pm 0.045)\%$  \\
Nominal + Veto Hits $< 2$&$(0.303 \pm 0.074 \pm 0.045)\%$ \\
Nominal + $-1.5<\chi_{tot}^\prime<0$ & $(0.304 \pm 0.077 \pm 0.045)\%$ \\
Nominal + $D>50$ cm \& $Y>-50$ cm& $(0.252 \pm 0.071 \pm 0.045)\%$      \\
Nominal + $D>75$ cm & $(0.222 \pm 0.074 \pm 0.045)\%$      \\
Nominal + $Y>-120$ cm& $(0.239 \pm 0.061 \pm 0.045)\%$      \\
Nominal + $\Delta t_{past}>15.2\mu$s \& $Y>-120$ cm& $(0.193 \pm 0.055 \pm 0.045)\%$      \\
Nominal + $S > 0.5$ &$(0.293 \pm 0.069 \pm 0.045)\%$ \\
\hline
\end{tabular}
\label{tab:oscillation_probs}
\end{table}
\newpage

\begin{table*}[\taboptions]

\caption{Number of beam on-off excess events that satisfy the selection
criteria for the primary $\bar \nu_\mu \rightarrow \bar \nu_e$
oscillation search with $1$ associated $\gamma$ and
with $>1$ associated $\gamma$. (An associated $\gamma$
is defined to have $R_\gamma>10$.)
The excess of events with
$>1$ correlated $\gamma$ is approximately zero, which is what is
expected for the reaction $\bar \nu_e p \rightarrow e^+ n$.}

\vspace{10mm}

\begin{tabular}{|c|c|c|}
\hline
Energy Selection&1 Associated $\gamma$&$>1$ Associated $\gamma$ \\
\hline
$20<E_e<60$ MeV&$49.1\pm9.4$&$-2.8\pm 2.4$ \\
$36<E_e<60$ MeV&$28.3\pm6.6$&$-3.0\pm 1.7$ \\
\hline
\end{tabular}
\label{tab:gammas}
\end{table*}

\clearpage
\begin{table*}[\taboptions]

\caption{Number of beam on-off excess events that satisfy the selection
criteria for the primary $\bar \nu_\mu \rightarrow \bar \nu_e$
oscillation search with $36<E_e<60$ MeV and with
$>11$ ``lookback'' hits in the 0-3 $\mu$s and
3-6 $\mu$s intervals. Results are shown for events with $R_\gamma \ge 0$
and for events with $R_\gamma >10$. The number of excess events 
in each $3 \mu$s interval is
consistent with the probability of having an accidental lookback
in the time interval.}

\vspace{10mm}

\begin{tabular}{|c|c|c|c|}
\hline
$R_\gamma$ Selection&$0-3 \mu$s&$3-6 \mu$s&Events Expected Due to Accidentals \\
\hline
$R_\gamma \ge 0$ &$11.5 \pm 6.3$&$7.8\pm 5.9$&$10.8\pm 2.2$ \\
$R_\gamma > 10$ &$1.7 \pm 1.4$&$0.5\pm 1.0$&$1.6\pm 0.4$ \\
\hline
\end{tabular}
\label{tab:look}
\end{table*}

%
\begin{table*}[\taboptions]
\caption{The number of excess events in the $60<E_e<200$ MeV
energy range, together with the corresponding oscillation
probability if the excess is due to 
$\nu_\mu \rightarrow \nu_e$ oscillations.
Also shown are the results from the higher precision analysis of
the 1993-1995 data sample \cite{paper3}.
}
\vspace{10mm}
\begin{tabular}{|c|c|c|}
\hline
Analysis& Excess Events&Oscillation Probability \\
\hline
Present Analysis (1993-1998)&$8.1 \pm 12.2 \pm 1.7$
&$(0.10\pm 0.16 \pm 0.04)\%$ \\
Previous Analysis (1993-1995)&$18.1 \pm 6.6 \pm 4.0$
&$(0.26 \pm 0.10 \pm 0.05)\%$ \\
\hline
\end{tabular}
\label{tab:table_he}
\end{table*}

\begin{table*}[\taboptions]
\caption{  The eight contributions to the $(\sin^22\theta,\Delta m^2)$
$\mathcal{L}$ fit from all
of the signal and background processes. Also shown are the fitted
number of events at the best fit point of
$(\sin^22\theta, \Delta m^2)_{best-fit} = (0.003,1.2 \rm{eV}^2)$.
        }
\vspace{10mm}
\begin{tabular}{|c|c|c|c|}
\hline
$\mathcal{L}$ Contribution & Signal or Background Source & Process &
Fitted Number of Events \\ \hline
1 &
$\overline\nu_\mu\rightarrow\overline\nu_e$ &
$\bar \nu_e p \rightarrow e+ n$& 89.5 \\
2 & BUB&&3664.6 \\
3 & DAR $\nu_e$ &
$\nu_e\;^{12}\,C\rightarrow e^- N_{g.s.}$& 1865.0 \\
&& $\nu_e\;^{12}\,C\rightarrow e^- N^*$& \\
&& $\nu_e\;^{13}\,C \rightarrow e^- N$& \\
&& $\nu e \rightarrow \nu e$& \\
4 & DIF $\nu_\mu$ &
$\nu_\mu C\rightarrow \mu^- N^*$& 37.3 \\
&&  $\nu_\mu C\rightarrow \mu^- N_{g.s.}$& \\
5 & DIF $\overline\nu_\mu$ &
$\overline\nu_\mu p \rightarrow \mu^+ n$&5.9 \\
&&  $\overline\nu_\mu C \rightarrow \mu^+ B^*$& \\
&&  $\overline\nu_\mu C\rightarrow \mu^+ B_{g.s.}$& \\
6 & DAR $\overline \nu_e$ ($\mu^-$ DAR)&
$\overline\nu_e p \rightarrow e^+ n$&16.7 \\
7 &
$\nu_\mu \rightarrow\nu_e$&$\nu_e C \rightarrow e^- N$&6.1 \\
8 & DIF $\pi^+\rightarrow\nu_e$ and $\mu^+\rightarrow\nu_e$ decay &
$\nu_e C \rightarrow e^- N$&11.9 \\
\hline
\end{tabular}
\label{tab:tb2}
\end{table*}

\begin{table*}[\taboptions]
\caption{ A comparison of the LSND and KARMEN experiments.}
\vspace{10mm}
\begin{tabular}{|c|c|c|}
\hline
Property&LSND&KARMEN \\
\hline
Proton Energy&798 MeV&800 MeV \\
Proton Intensity&1000 $\mu$A&200 $\mu$A \\
Duty Factor&$6 \times 10^{-2}$&$1 \times 10^{-5}$ \\
Total Mass&167 t&56 t \\
Neutrino Distance&30 m&17.5 m \\
Particle Identification&YES&NO \\
Energy Resolution at 50 MeV&$6.6\%$&$1.6\%$ \\
\hline
\end{tabular}
\label{tab:LSNDKAR}
\end{table*}

\clearpage
\newcommand{\figscale}{0.7}
\newcommand{\figoptions}{tbp}

%
\begin{figure*}[\figoptions]
\vspace*{\fill}
\centering
\scalebox{\figscale}{
\includegraphics{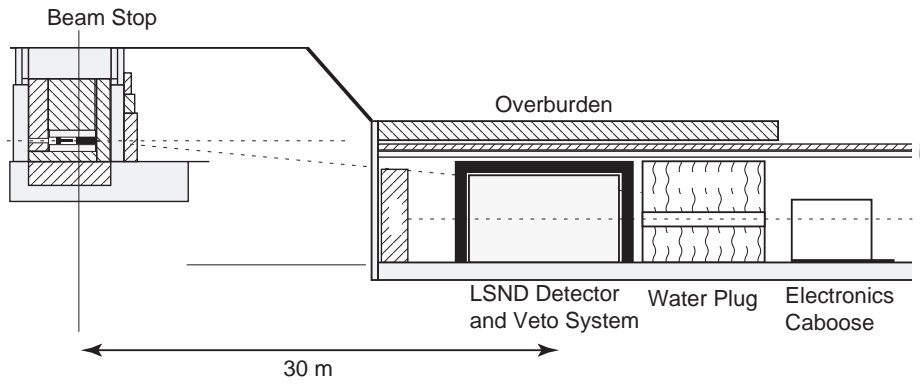}
}
\vspace{5mm}
\caption{
The layout of the LSND detector and the A6 beam stop area.
}
\label{fig:Fig{1}_detector_pic}
\end{figure*}

%
\begin{figure*}[\figoptions]
\vspace*{\fill}
\centering
\scalebox{\figscale}{
\includegraphics{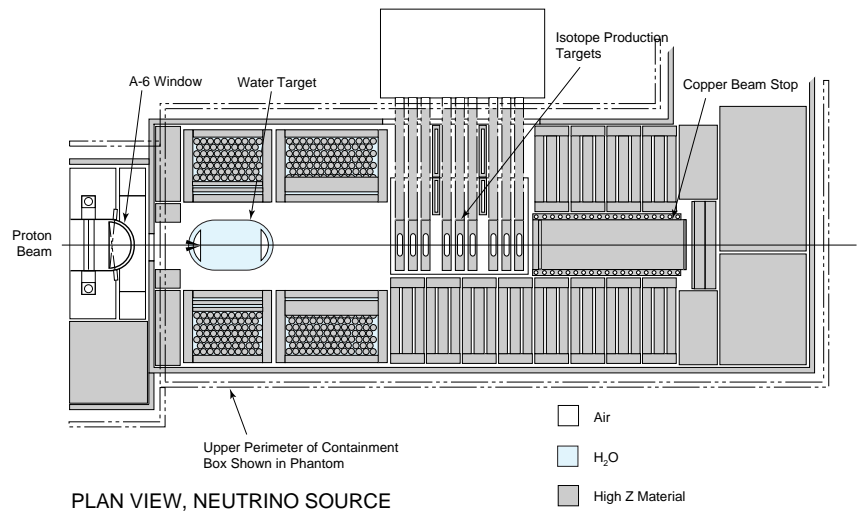}
}
\vspace{5mm}
\caption{
The layout of the A6 beam stop, as it was configured
for the 1993-1995 data taking.
}
\label{fig:Fig{2}_targets2}
\end{figure*}

%
\begin{figure*}[\figoptions]
\vspace*{\fill}
\centering
\scalebox{\figscale}{
\includegraphics{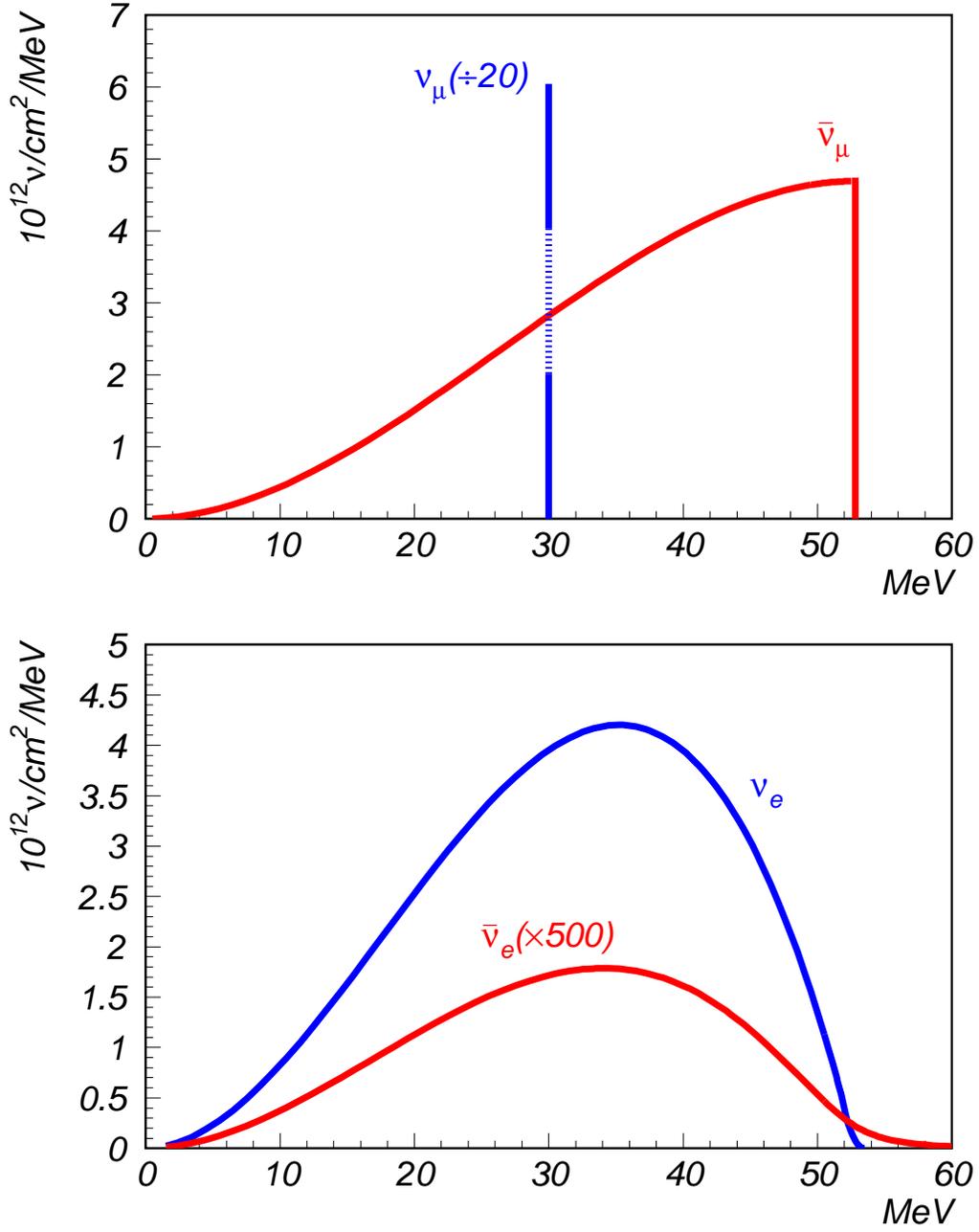}
}
\vspace{5mm}
\caption{
The decay-at-rest neutrino fluxes averaged over
the detector.
}
\label{fig:Fig{3}_dar_flux}
\end{figure*}

%
\begin{figure*}[\figoptions]
\vspace*{\fill}
\centering
\scalebox{\figscale}{
\includegraphics{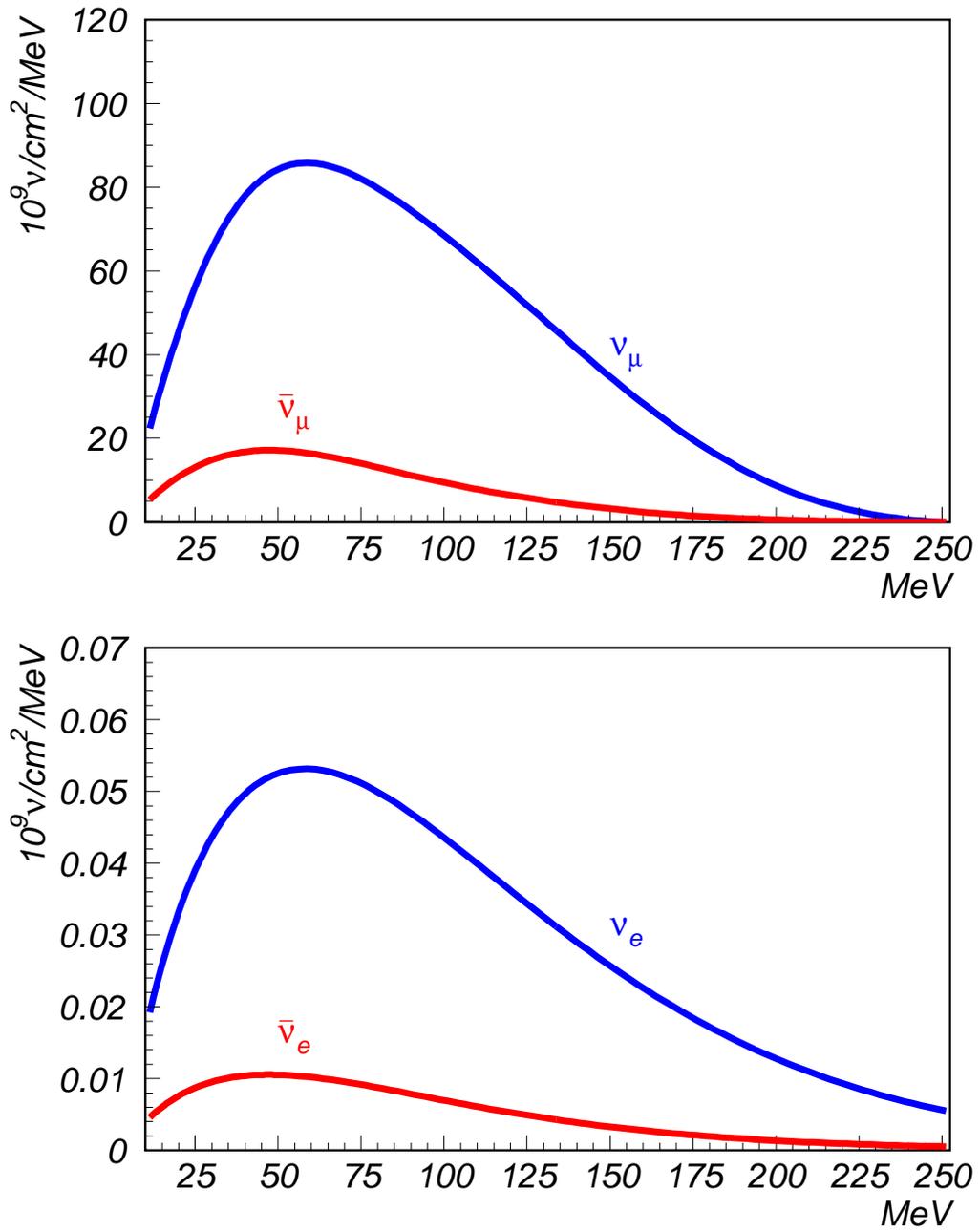}
}
\vspace{5mm}
\caption{
The decay-in-flight neutrino fluxes 
averaged over the detector.
}
\label{fig:Fig{4}_dif_flux}
\end{figure*}

%
%
\begin{figure*}[\figoptions]
\vspace*{\fill}
\centering
\scalebox{\figscale}{
\includegraphics{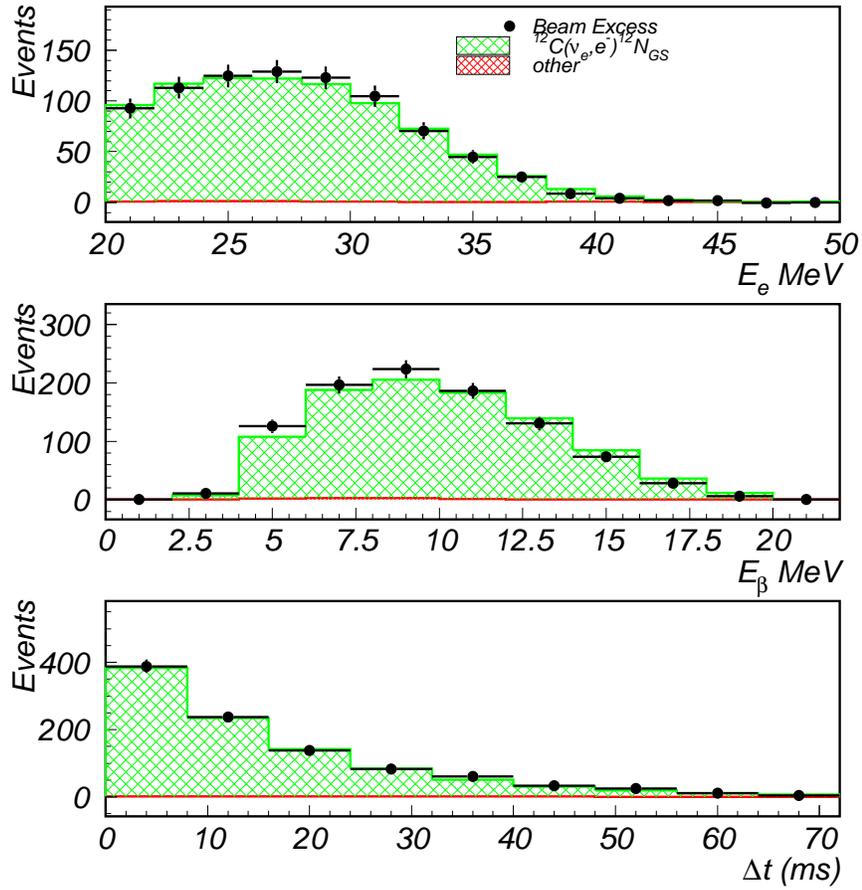}
}
\vspace{5mm}
\caption{
The electron and $\beta$ energy
distributions and the time between the electron and $\beta$, $\Delta t$,
for $^{12}C(\nu_e,e^-)^{12}N_{g.s.}$ scattering events.
}
\label{fig:Fig{5}_nuecgs_energy}
\end{figure*}

%
%
\begin{figure*}[\figoptions]
\vspace*{\fill}
\centering
\scalebox{\figscale}{
\includegraphics{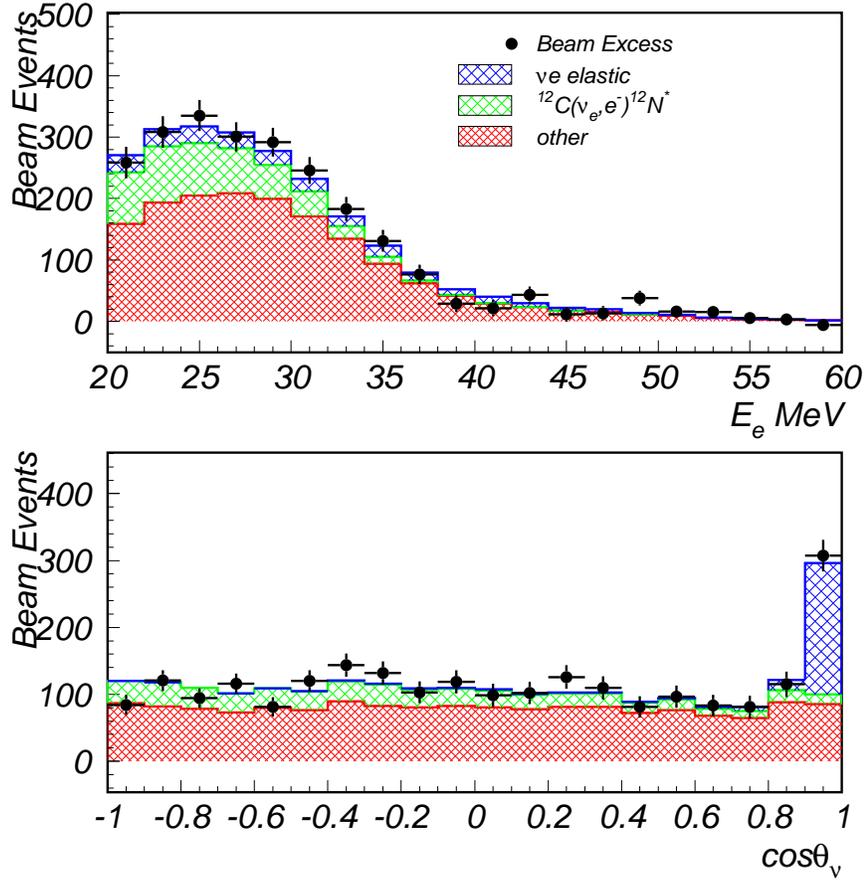}
}
\vspace{5mm}
\caption{
The energy and angular distributions for inclusive electron events.
$E_e$ is the electron
energy and $\theta_\nu$ is the angle between the incident neutrino and
outgoing electron directions. Neutrino-electron elastic scattering
events are clearly seen near $\cos \theta_\nu \sim 1$.
}
\label{fig:Fig{6}_elec_energy}
\end{figure*}

%
%
\begin{figure*}[\figoptions]
\vspace*{\fill}
\centering
\scalebox{\figscale}{
\includegraphics{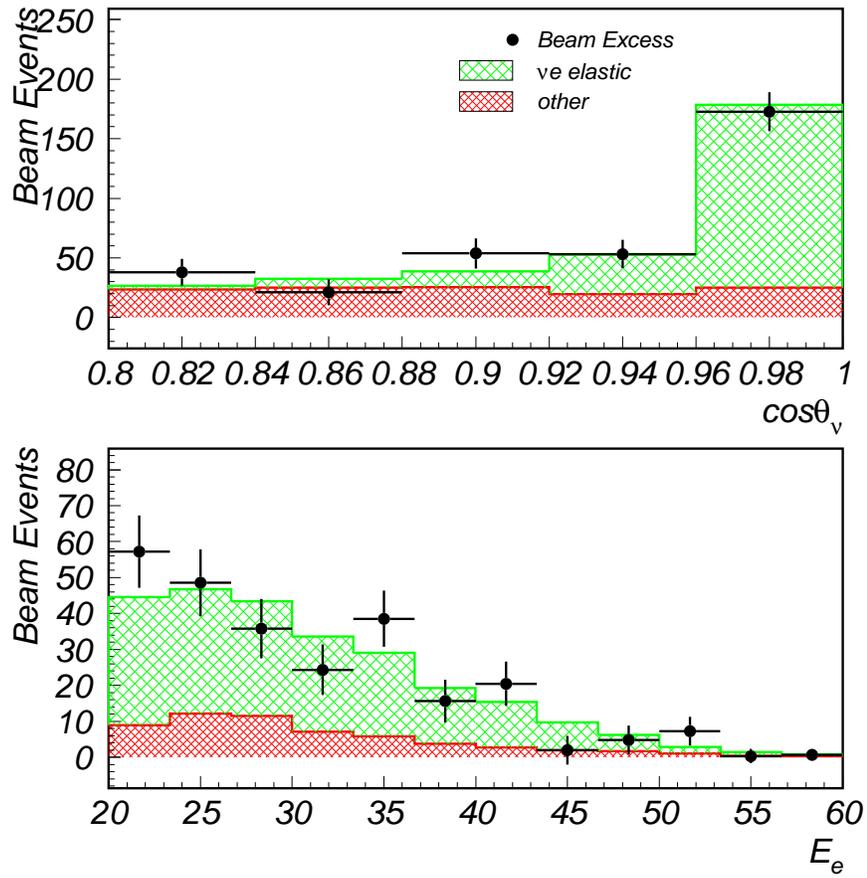}
}
\vspace{5mm}
\caption{
The angular distribution (top plot) and the
energy distribution (bottom plot) for neutrino-electron elastic scattering
events with $\cos \theta_\nu > 0.9$ and with
$^{12}C(\nu_e,e^-)^{12}N_{g.s.}$ events removed.
}
\label{fig:Fig{7}_elec_nuel}
\end{figure*}

%
%
\begin{figure*}[\figoptions]
\vspace*{\fill}
\centering
\scalebox{\figscale}{
\includegraphics{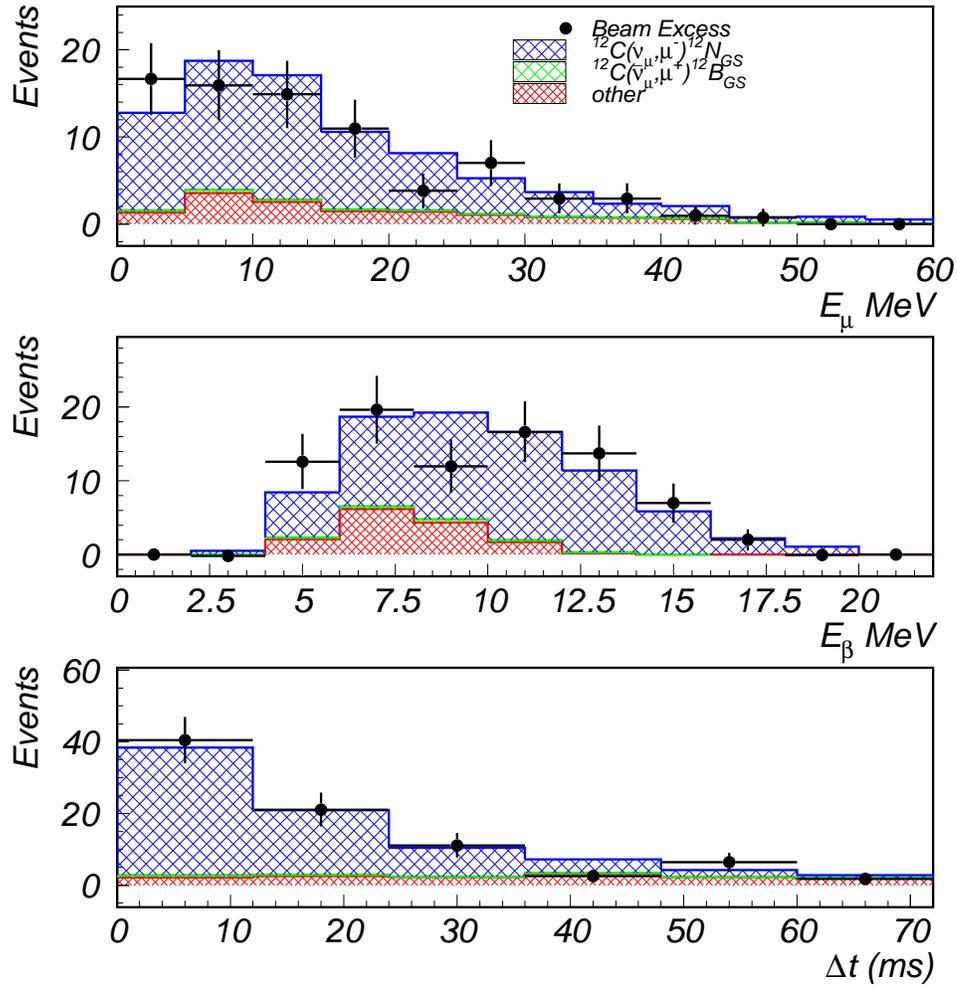}
}
\vspace{5mm}
\caption{
The muon and $\beta$ energy
distributions (electron energy equivalent)
and the time between the muon and $\beta$ for
$^{12}C(\nu_\mu,\mu^-)^{12}N_{g.s.}$ scattering events.
}
\label{fig:Fig{8}_numugs_energy}
\end{figure*}
\clearpage

%
%
\begin{figure*}[\figoptions]
\vspace*{\fill}
\centering
\scalebox{\figscale}{
\includegraphics{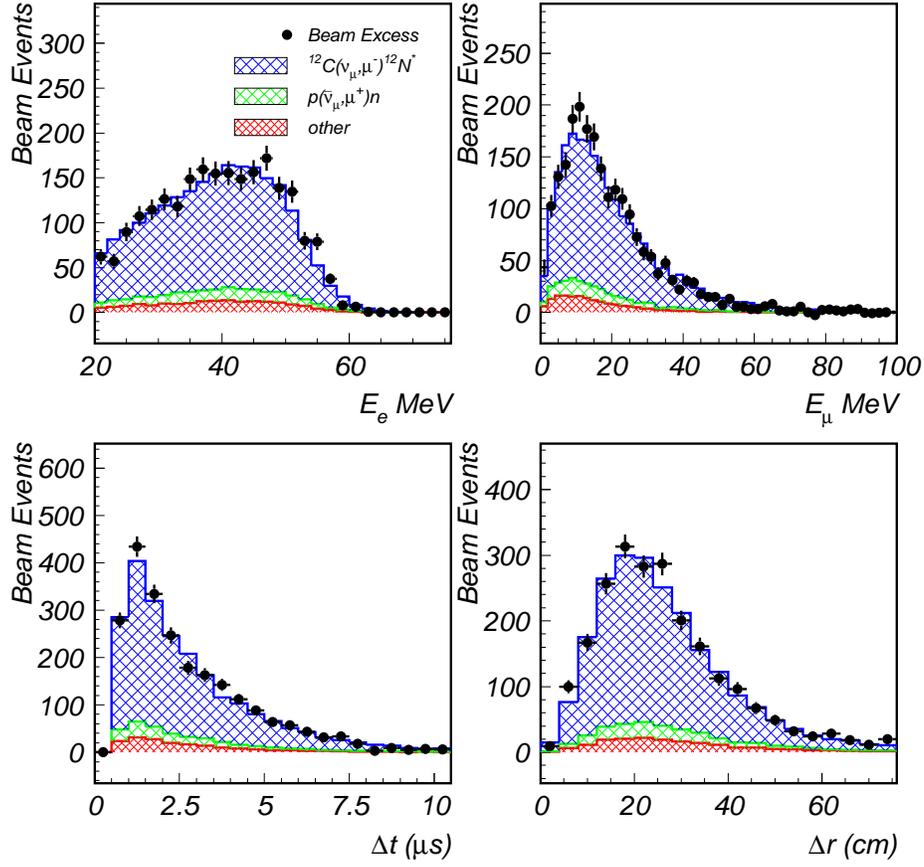}
}
\vspace{5mm}
\caption{
The Michel electron and muon
energy distributions (electron energy equivalent), 
the time between the muon and electron, $\Delta t$,
and the distance between the reconstructed electron position and muon position,
$\Delta r$, for $\nu_\mu C \rightarrow \mu^- N$,
$\bar \nu_\mu C \rightarrow \mu^+ B$, and
$\bar \nu_\mu p \rightarrow \mu^+ n$ inclusive
scattering events. 
}
\label{fig:Fig{9}_numu_energy}
\end{figure*}

%
%
\begin{figure*}[\figoptions]
\vspace*{\fill}
\centering
\scalebox{\figscale}{
\includegraphics{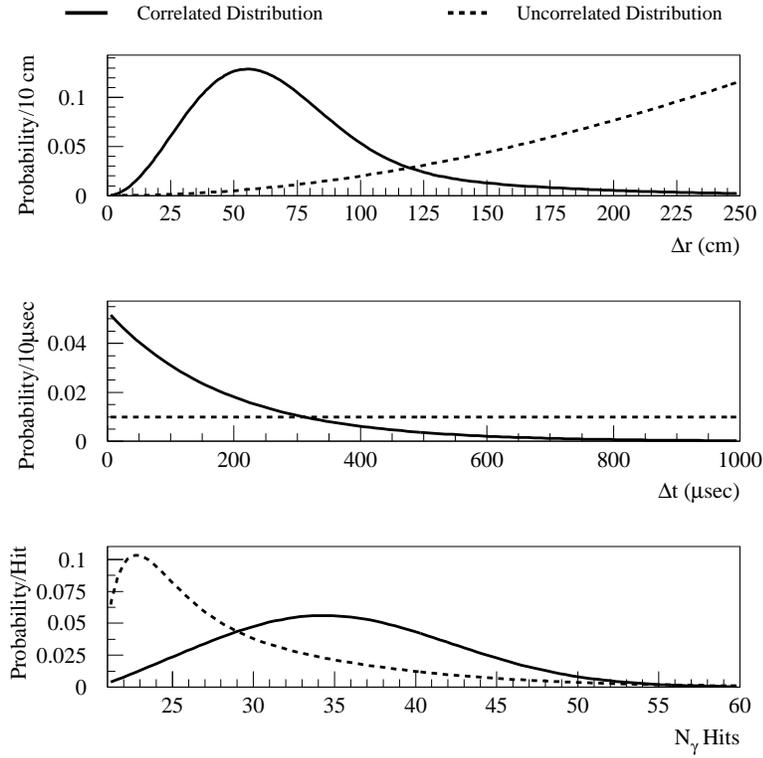}
}
\vspace{5mm}
\caption{
Distributions for correlated 2.2 MeV $\gamma$ (solid curves)
and accidental $\gamma$ (dashed curves). The top plot shows
the distance between the reconstructed
$\gamma$ position and positron position, $\Delta r$, the middle plot shows
the time interval between the $\gamma$ and
positron, $\Delta t$, and the bottom plot shows the number of hit phototubes
associated with the $\gamma$, $N_{hits}$.
}
\label{fig:Fig{10}_ruld}
\end{figure*}

%
%
\begin{figure*}[\figoptions]
\vspace*{\fill}
\centering
\scalebox{\figscale}{
\includegraphics{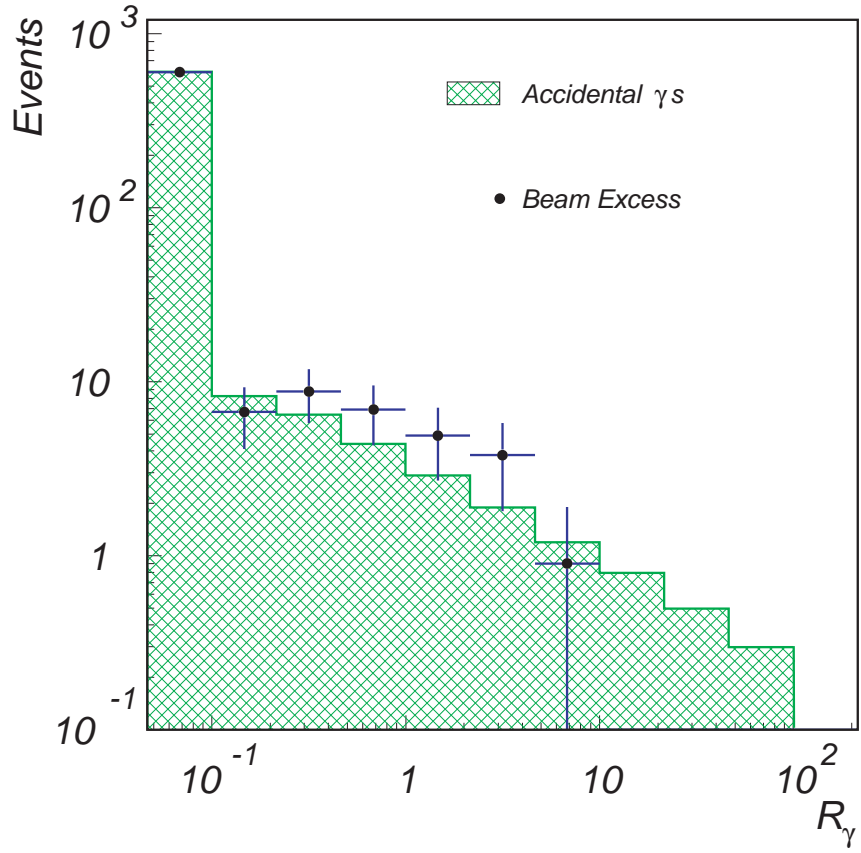}
}
\vspace{5mm}
\caption{
The $R_\gamma$ distribution for $\nu_e C \rightarrow e^- N_{g.s.}$
exclusive events, where
the $N_{g.s.}$ $\beta$ decays. The distribution is consistent with
a pure accidental $\gamma$ shape.
}
\label{fig:Fig{11}_r2_nuec}
\end{figure*}

%
%
\begin{figure*}[\figoptions]
\vspace*{\fill}
\centering
\scalebox{\figscale}{
\includegraphics{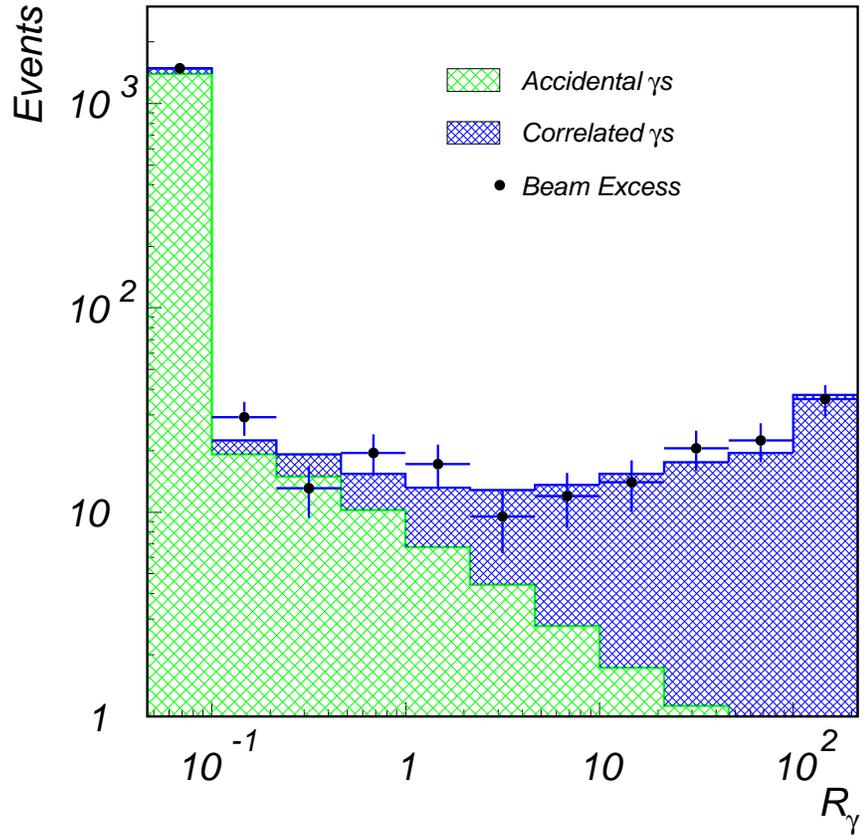}
}
\vspace{5mm}
\caption{
The $R_\gamma$ distribution for $\nu_\mu C \rightarrow \mu^- N$,
$\bar \nu_\mu C \rightarrow \mu^+ B$, and
$\bar \nu_\mu p \rightarrow \mu^+ n$ inclusive
scattering events.
}
\label{fig:Fig{12}_r2_numu}
\end{figure*}

%
%
\begin{figure*}[\figoptions]
\vspace*{\fill}
\centering
\scalebox{\figscale}{
\includegraphics{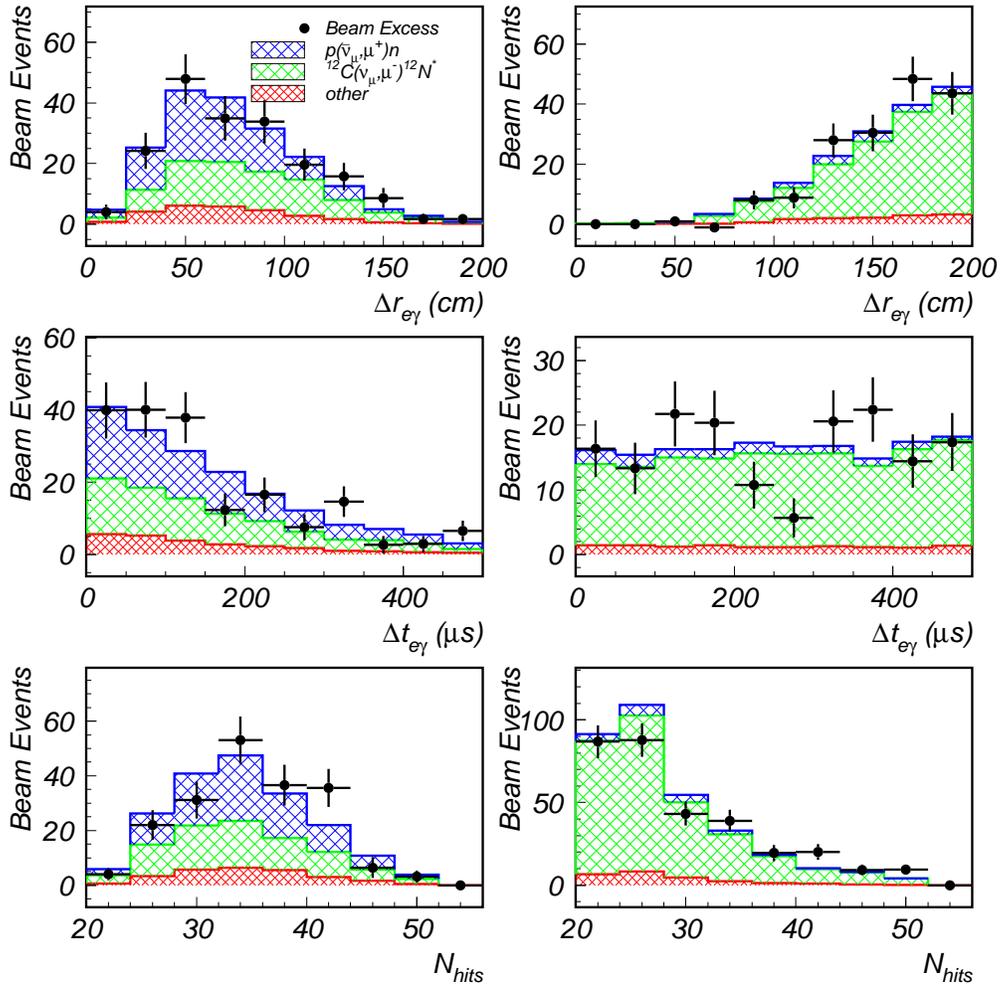}
}
\vspace{5mm}
\caption{
The individual $\gamma$ distributions
from $\nu_\mu C \rightarrow \mu^- N$, $\bar \nu_\mu C \rightarrow \mu^+ B$,
and $\bar \nu_\mu p \rightarrow \mu^+ n$ scattering for events with $R_\gamma >1$
(left side) and $R_\gamma <1$ (right side). The top plots show
the distance between the reconstructed
$\gamma$ position and positron position, $\Delta r$, the middle plots show
the time interval between the $\gamma$ and
positron, $\Delta t$, and the bottom plots show the number of hit phototubes
associated with the $\gamma$, $N_{hits}$.
}
\label{fig:Fig{13}_numu_gamma}
\end{figure*}

%
%
\begin{figure*}[\figoptions]
\vspace*{\fill}
\centering
\scalebox{\figscale}{
\includegraphics{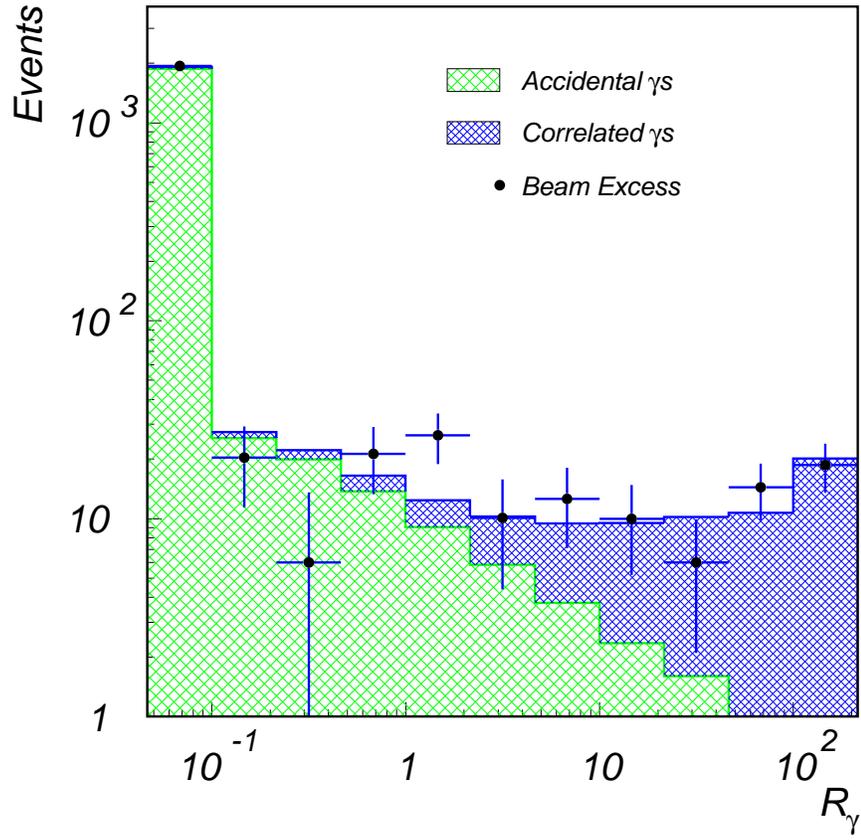}
}
\vspace{5mm}
\caption{
The $R_\gamma$ distribution for events that satisfy the selection
criteria for the primary $\bar \nu_\mu \rightarrow \bar \nu_e$
oscillation search.
}
\label{fig:Fig{14}_r2_osc}
\end{figure*}

%
%
\begin{figure*}[\figoptions]
\vspace*{\fill}
\centering
\scalebox{\figscale}{
\includegraphics{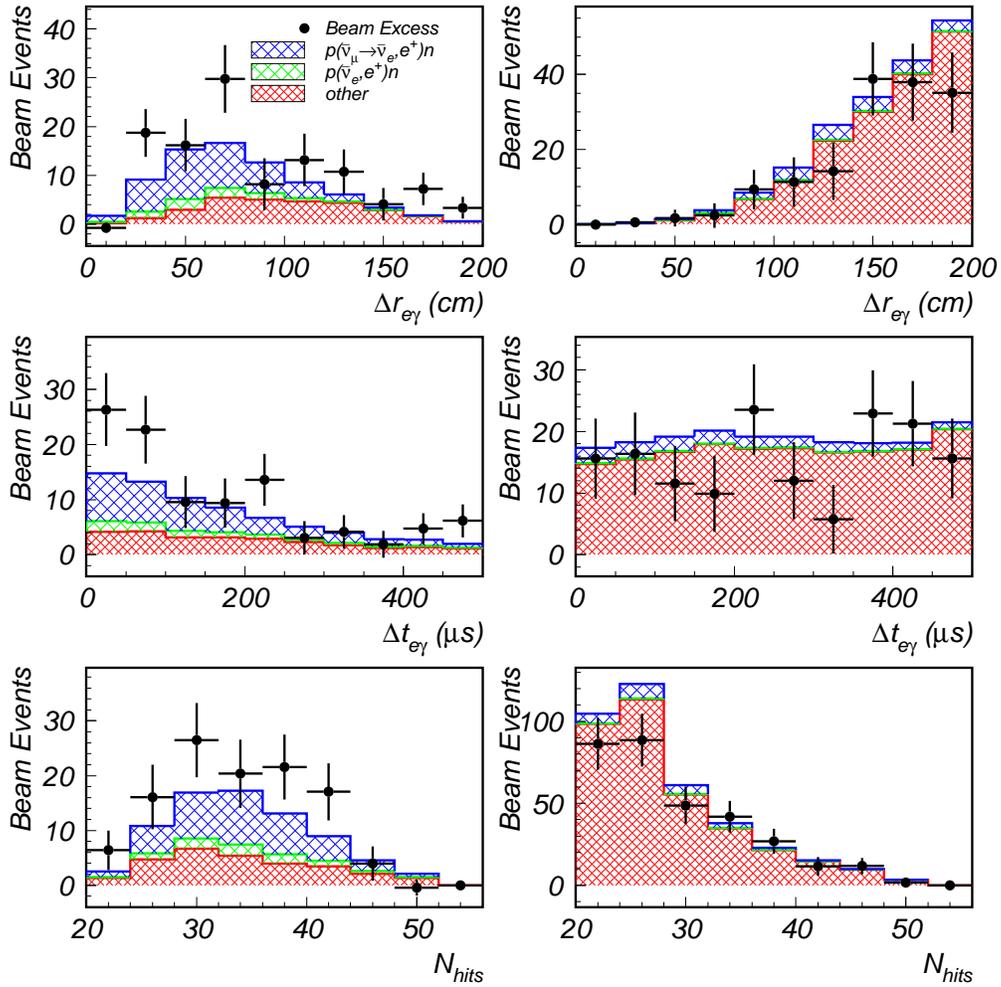}
}
\vspace{5mm}
\caption{
The individual $\gamma$ distributions
for events that satisfy the selection
criteria for the primary $\bar \nu_\mu \rightarrow \bar \nu_e$
oscillation search with $R_\gamma >1$
(left side) and $R_\gamma <1$ (right side).
The top plots show
the distance between the reconstructed
$\gamma$ position and positron position, $\Delta r$, the middle plots show
the time interval between the $\gamma$ and
positron, $\Delta t$, and the bottom plots show the number of hit phototubes
associated with the $\gamma$, $N_{hits}$.
}
\label{fig:Fig{15}_elec_gamma}
\end{figure*}

%
%
\begin{figure*}[\figoptions]
\vspace*{\fill}
\centering
\scalebox{\figscale}{
\includegraphics{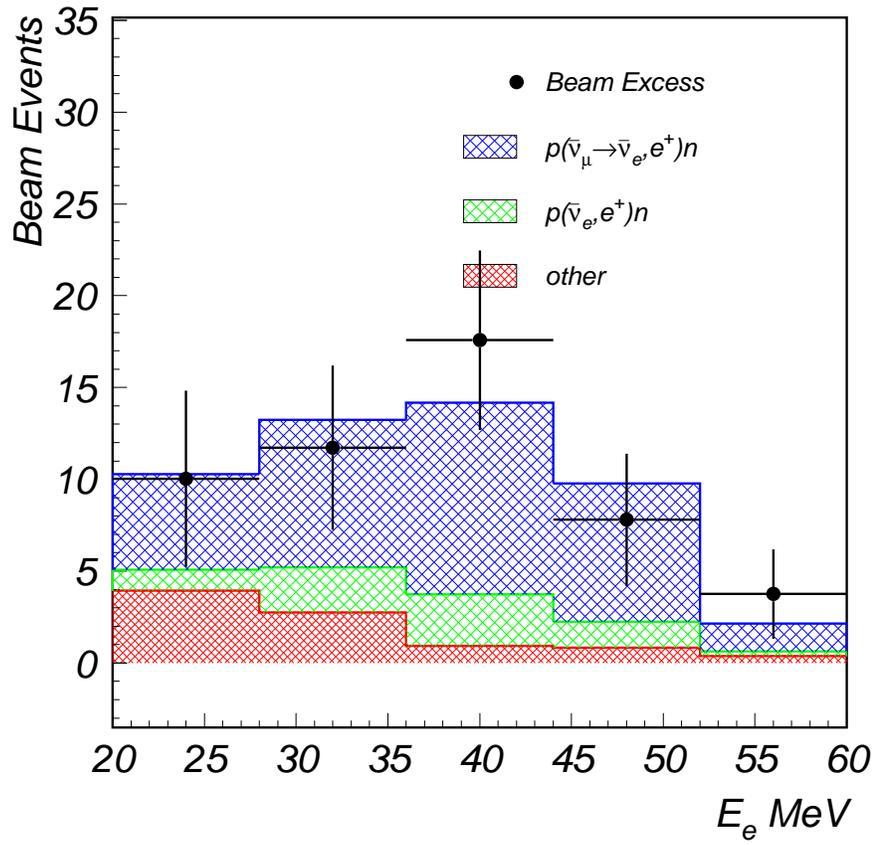}
}
\vspace{5mm}
\caption{
The energy distribution of
the 1993-1998 data sample for events with $R_\gamma >10$. 
The shaded region shows the
expected distribution
from a combination of neutrino background plus neutrino oscillations
at low $\Delta m^2$.
}
\label{fig:Fig{16}_elec_en_rcut}
\end{figure*}

%
%
\begin{figure*}[\figoptions]
\vspace*{\fill}
\centering
\scalebox{\figscale}{
\includegraphics{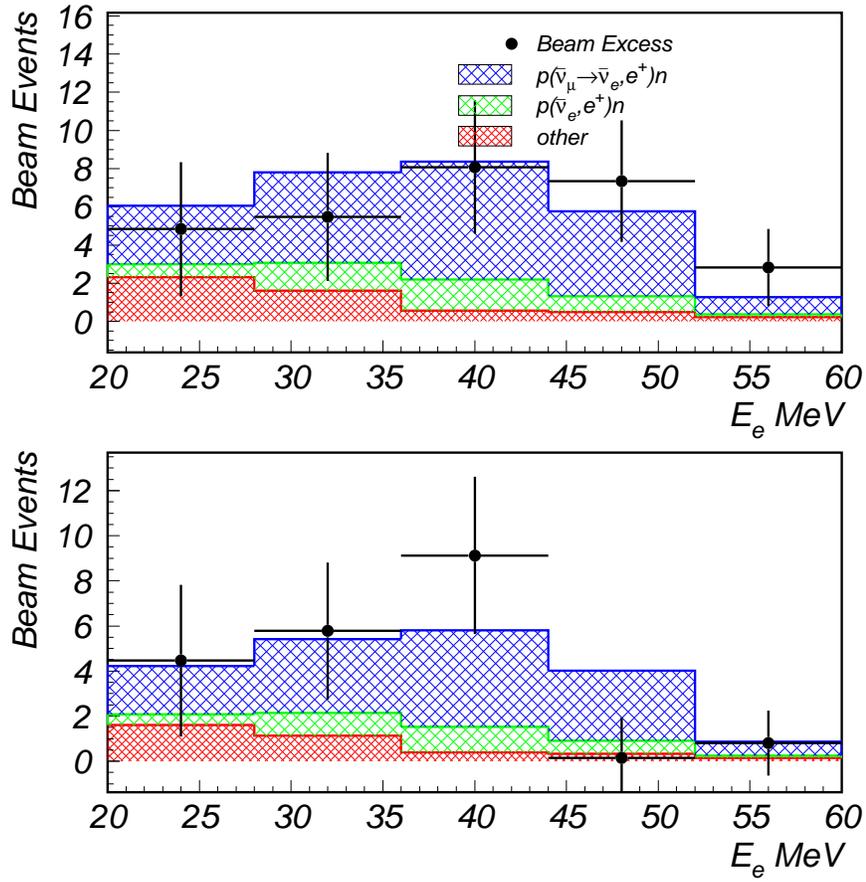}
}
\vspace{5mm}
\caption{
The energy distribution of
the 1993-1995 (top plot) and 1996-1998 (bottom plot) 
data samples for events with $R_\gamma >10$. 
The shaded region shows the
expected distribution
from a combination of neutrino background plus neutrino oscillations
at low $\Delta m^2$.
}
\label{fig:Fig{17}_elec_en_rcut1}
\end{figure*}

%
%
\begin{figure*}[\figoptions]
\vspace*{\fill}
\centering
\scalebox{\figscale}{
\includegraphics{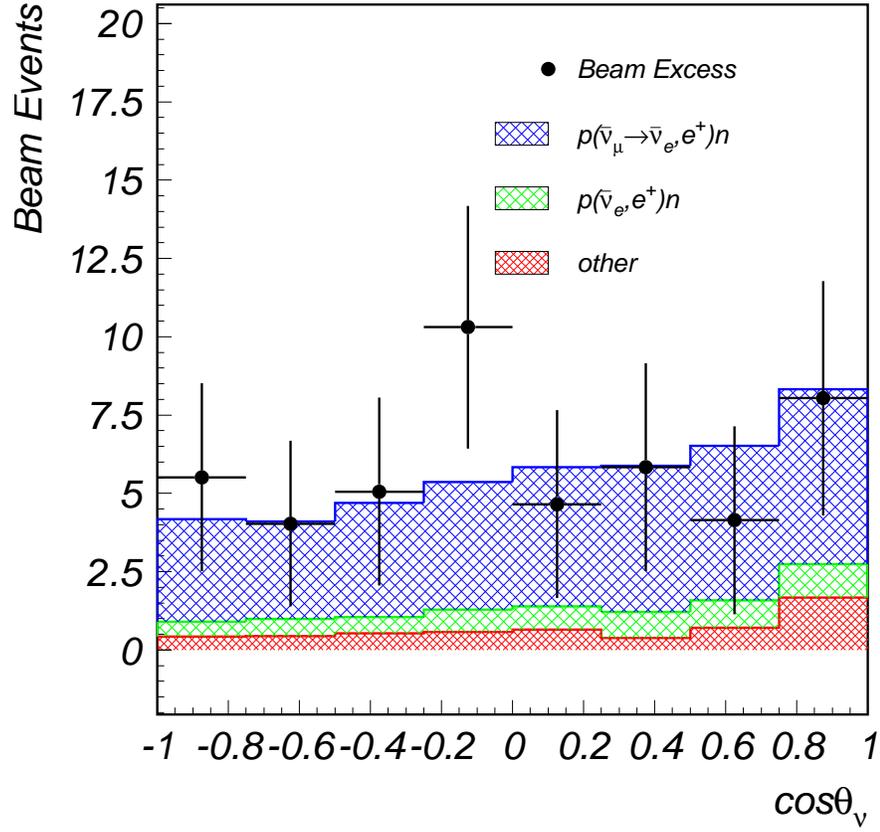}
}
\vspace{5mm}
\caption{
The $\cos\theta_\nu$ distribution for events with $R_\gamma>1$ 
and $36<E<60$ MeV. The shaded region shows the expected distribution
from a combination of neutrino background plus neutrino oscillations
at low $\Delta m^2$.
}
\label{fig:Fig{18}_elec_cos_rcut}
\end{figure*}

%
%
\begin{figure*}[\figoptions]
\vspace*{\fill}
\centering
\scalebox{\figscale}{
\includegraphics{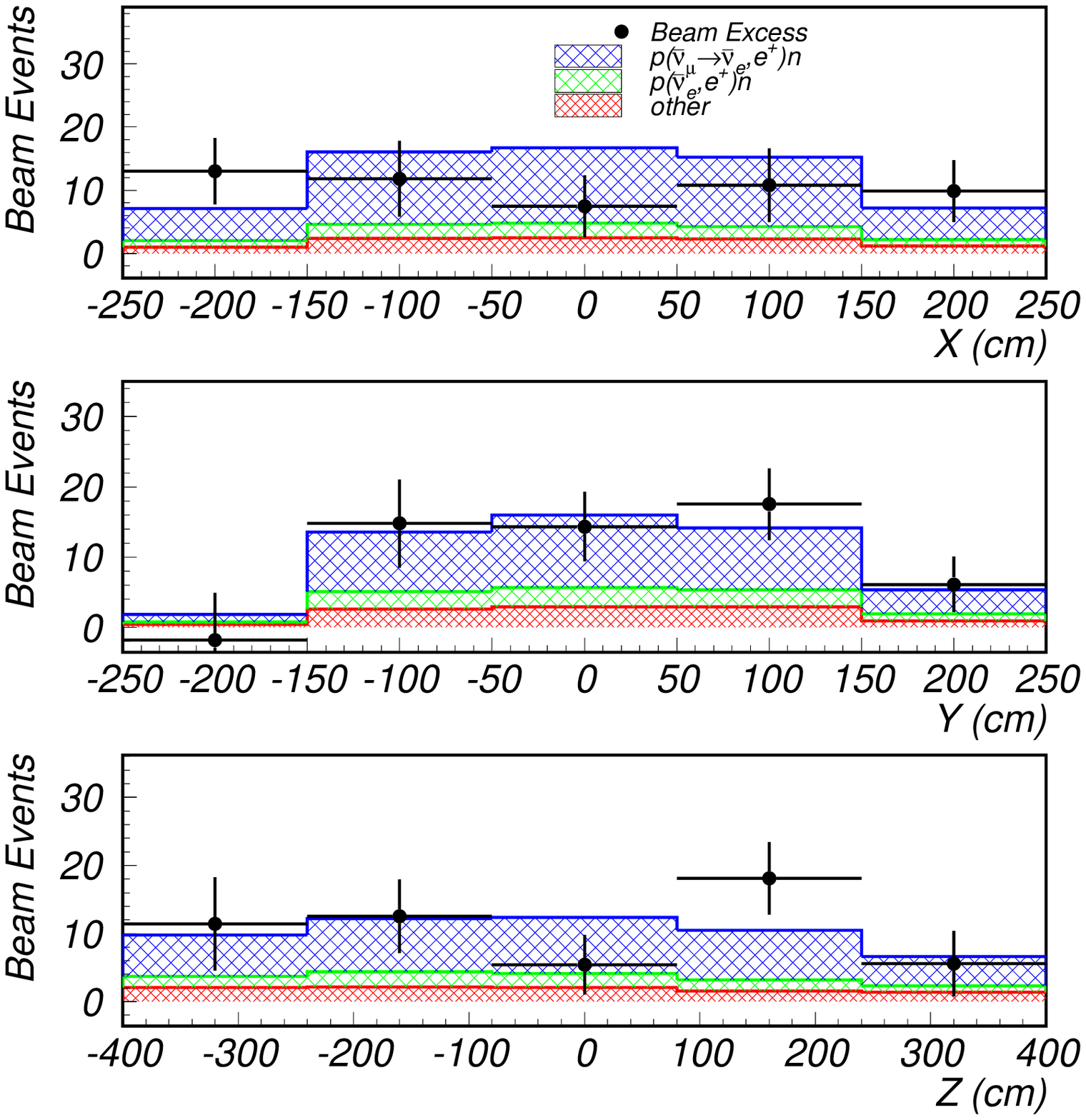}
}
\vspace{5mm}
\caption{
The spatial distributions for
events with $R_\gamma >10$, $20<E_e<60$ MeV, and $D > 10$ cm. 
The shaded region shows the 
expected distribution
from a combination of neutrino background plus neutrino oscillations
at low $\Delta m^2$.
}
\label{fig:Fig{19}_elec_space_rcut}
\end{figure*}

%
%
\begin{figure*}[\figoptions]
\vspace*{\fill}
\centering
\scalebox{\figscale}{
\includegraphics{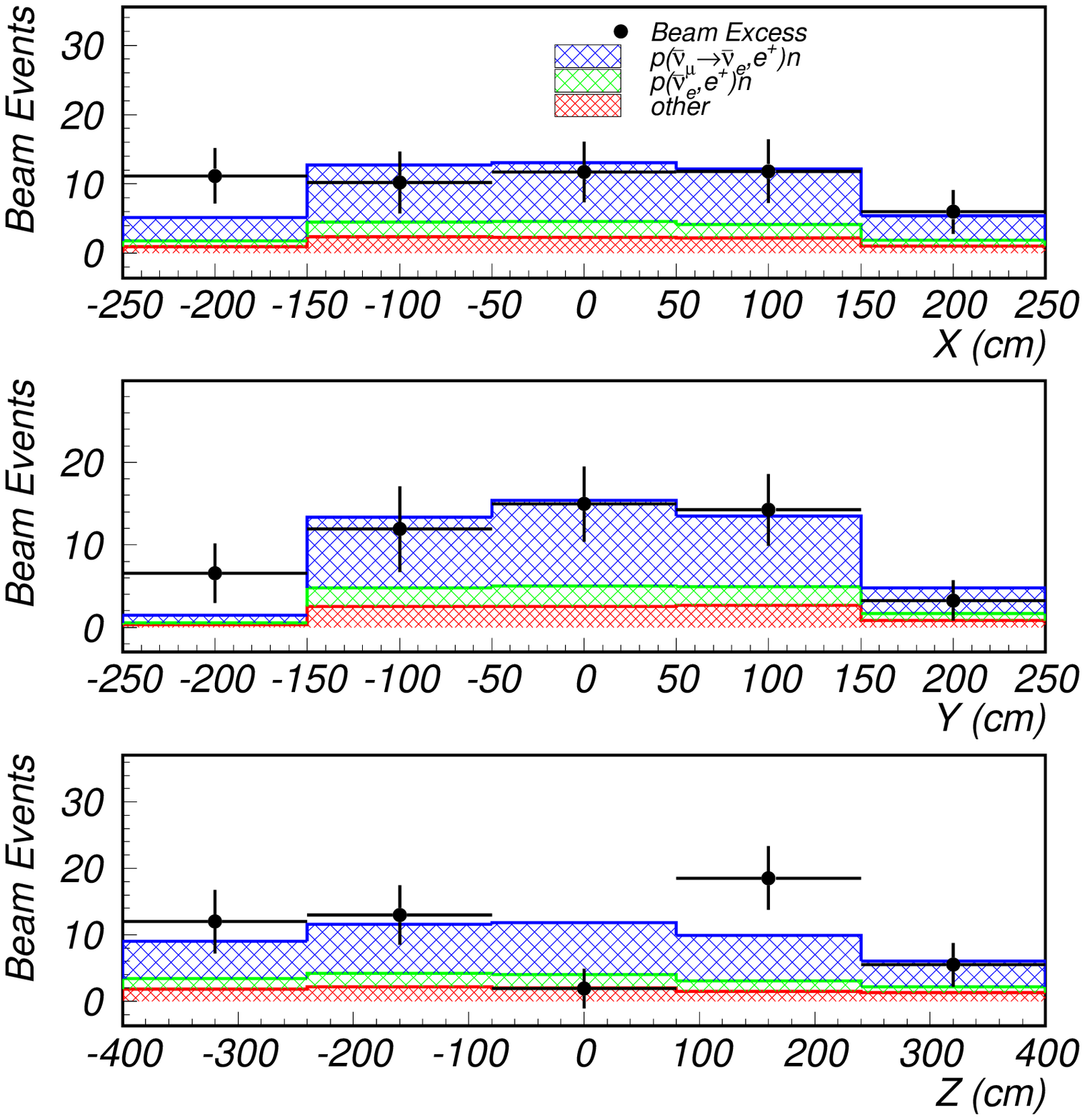}
}
\vspace{5mm}
\caption{
The spatial distributions for
events with $R_\gamma >10$, $20<E_e<60$ MeV, and $D > 35$ cm. 
The shaded region shows the 
expected distribution
from a combination of neutrino background plus neutrino oscillations
at low $\Delta m^2$.
}
\label{fig:Fig{20}_elec_space_rcut2}
\end{figure*}

%
%
\begin{figure*}[\figoptions]
\vspace*{\fill}
\centering
\scalebox{\figscale}{
\includegraphics{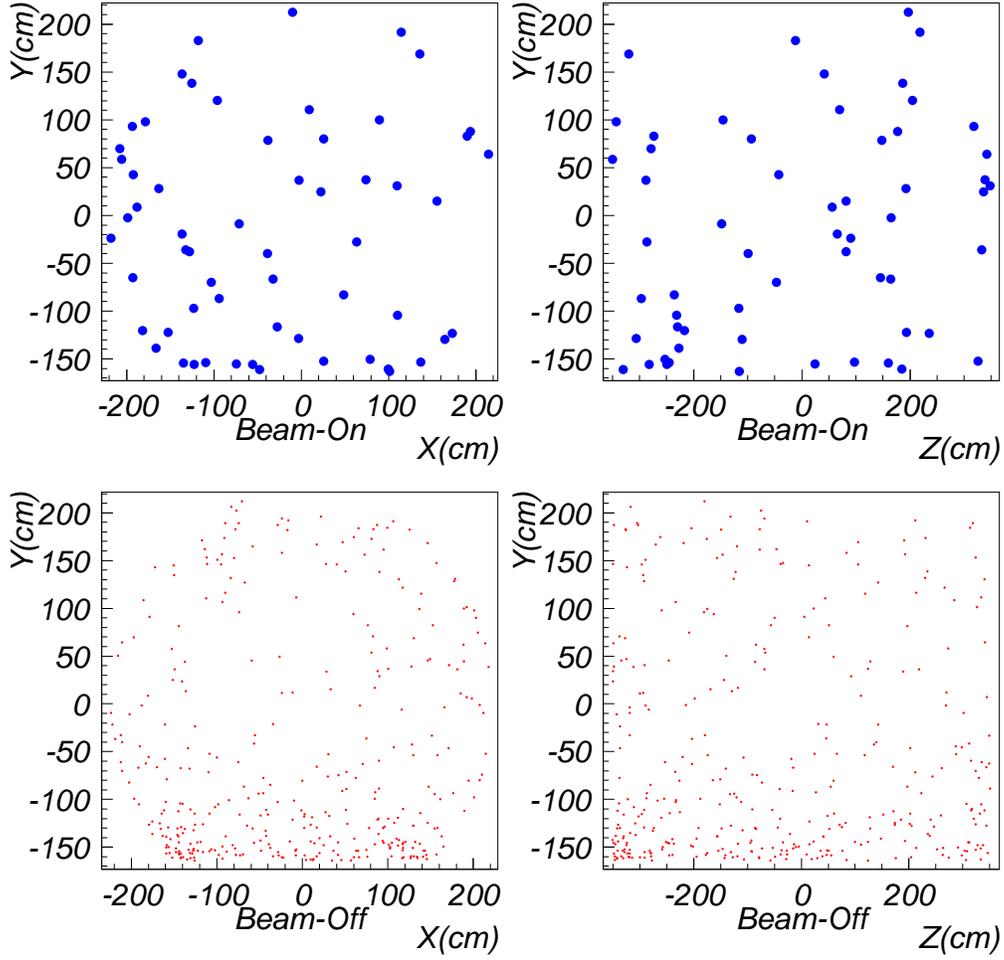}
}
\vspace{5mm}
\caption{
The scatter plots of the x-y and y-z spatial distributions for
events with $R_\gamma >10$, $20<E_e<60$ MeV, and $D > 35$ cm. 
Beam-on and beam-off
events are shown separately. The ratio of the dot area in 
beam-off plots, to the dot area in the beam-on plots, is equal
to the duty ratio. This gives the appropriate scale for the 
beam-off subtraction.
}
\label{fig:Fig{21}_scat_xyz}
\end{figure*}

%
%
\begin{figure*}[\figoptions]
\vspace*{\fill}
\centering
\scalebox{\figscale}{
\includegraphics{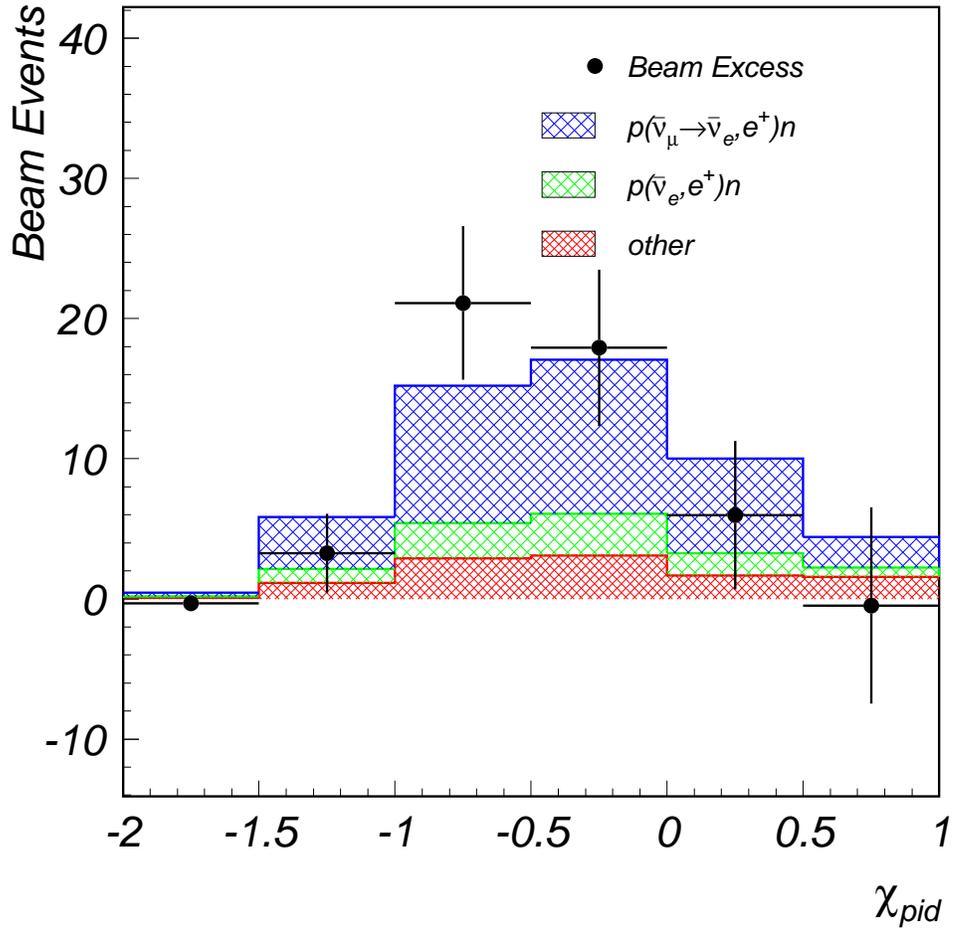}
}
\vspace{5mm}
\caption{
The particle identification, $\chi_{tot}^\prime$, distribution for
events with $R_\gamma >10$, $20<E_e<60$ MeV, and $D > 35$ cm. 
The shaded region shows the 
expected distribution
from a combination of neutrino background plus neutrino oscillations
at low $\Delta m^2$. Oscillation candidate events are required to
staisfy the requirement $-1.5 < \chi_{tot}^\prime < 0.5$.
}
\label{fig:Fig{22}_elec_pid_rcut}
\end{figure*}
\clearpage

%
%
\begin{figure*}[\figoptions]
\vspace*{\fill}
\centering
\scalebox{\figscale}{
\includegraphics{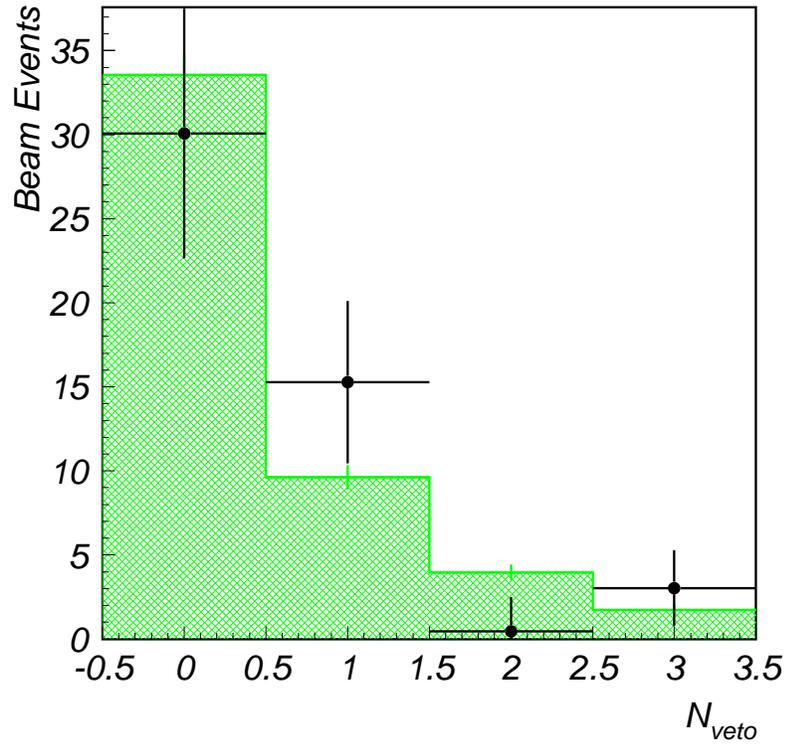}
}
\vspace{5mm}
\caption{
The veto hit distribution for
events with $R_\gamma >10$ and $20<E_e<60$ MeV. 
The data agree well with the distribution from
$\nu_e C \rightarrow e^- N_{g.s.}$ scattering (shaded histogram), where
the reaction is identified by the $ N_{g.s.}$ $\beta$ decay.
}
\label{fig:Fig{23}_elec_veto_rcut}
\end{figure*}

%
%
\begin{figure*}[\figoptions]
\vspace*{\fill}
\centering
\scalebox{\figscale}{
\includegraphics{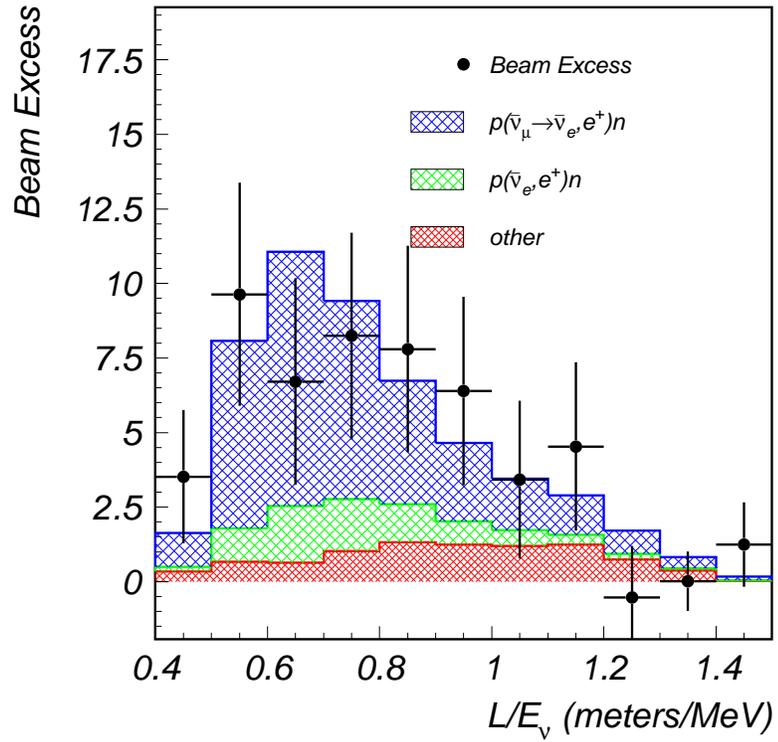}
}
\vspace{5mm}
\caption{
The $L_\nu/E_\nu$ distribution
for events with $R_\gamma >10$
and $20<E_e<60$ MeV, where $L_\nu$ is the distance travelled by the
neutrino in meters and $E_\nu$ is the neutrino energy in MeV. The
data agree well with the expectation from neutrino background and
neutrino oscillations at low $\Delta m^2$.
}
\label{fig:Fig{24}_elec_loe_rcut}
\end{figure*}

%
%
\begin{figure*}[\figoptions]
\vspace*{\fill}
\centering
\scalebox{\figscale}{
\includegraphics{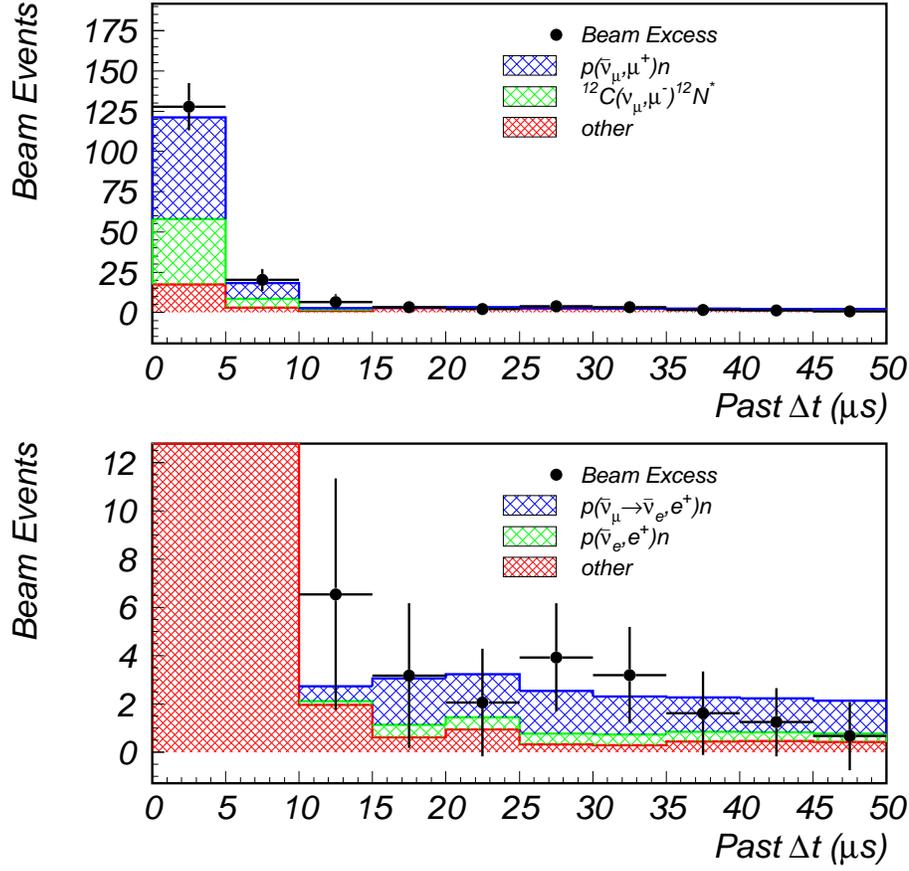}
}
\vspace{5mm}
\caption{
This figure shows the time to the previous event for $R_\gamma>10$
electron events prior to applying the $\Delta t_{past}>12\mu$s
selection. In the upper graph the beam excess events are in agreement
with our expectations for $\nmbp$ and $\nmbc$ processes. With the same
data on a smaller vertical scale, the bottom graph shows events with
accidental past activities, in agreement with expectations from random
cosmic ray backgrounds and beam related backgrounds. Note that most of
the oscillation candidate events have no past activity, and therefore
do not appear in these graphs.}
\label{fig:Fig{25}_elec_pdtmin2}
\end{figure*}

%
%
\begin{figure*}[\figoptions]
\vspace*{\fill}
\centering
\scalebox{0.7}{
\includegraphics{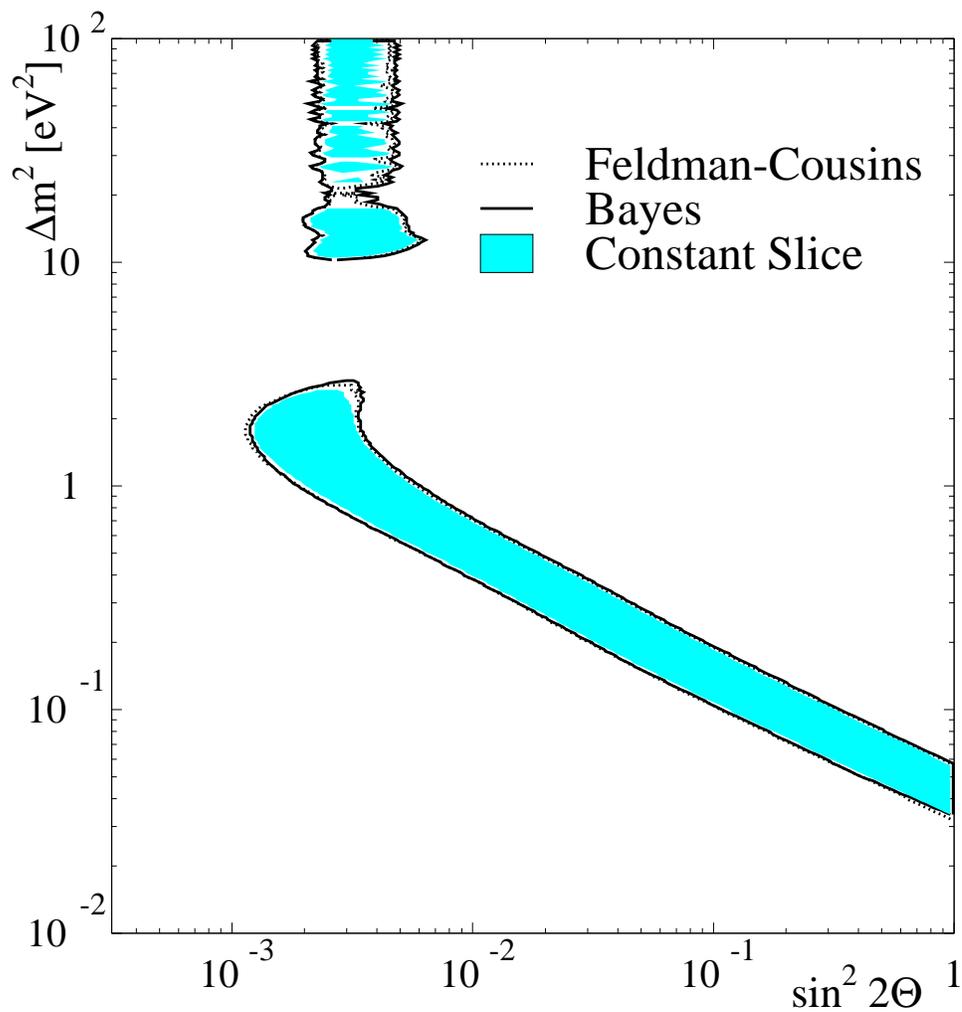}
}
\vspace{5mm}
\caption{
Favored regions in the $(\sin^22\theta,\Delta m^2)$ plane at $90\%$ CL.
The Feldman-Cousins, Bayesian,
and constant-slice
methods all give about the same result.
}
\label{fig:Fig{26}_cl90}
\end{figure*}

%
%
\begin{figure*}[\figoptions]
\vspace*{\fill}
\centering
\scalebox{\figscale}{
\includegraphics{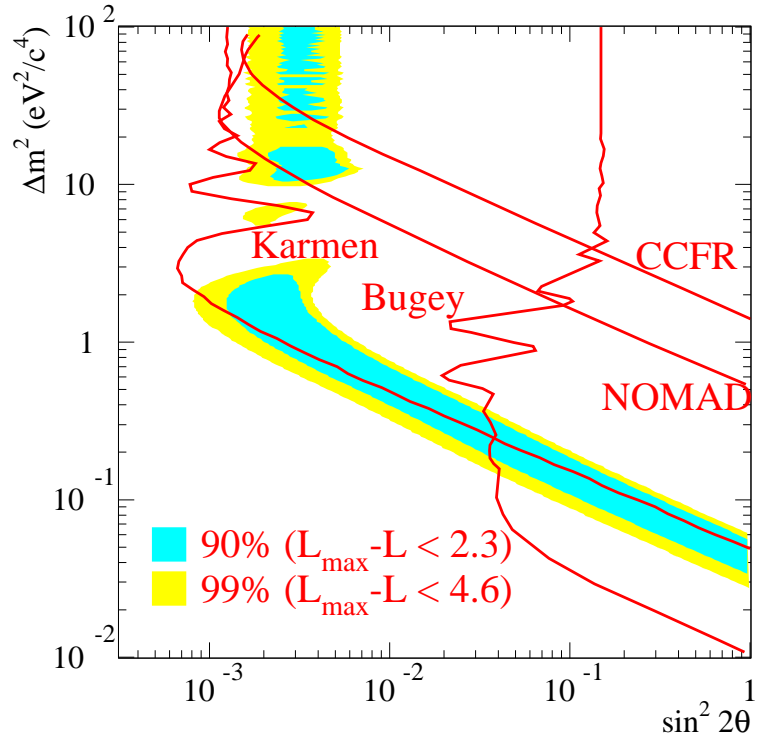}
}
\vspace{5mm}
\caption{
A $(\sin^22\theta,\Delta m^2)$
oscillation parameter fit
for the entire data sample,
$20<E_e<200$ MeV. The fit includes
primary $\bar \nu_\mu \rightarrow \bar \nu_e$
oscillations and secondary $\nu_\mu \rightarrow \nu_e$
oscillations, as well as all known neutrino backgrounds. The inner and
outer regions correspond to $90\%$ and $99\%$ CL allowed regions, while the
curves are $90\%$ CL limits from
the Bugey reactor experiment,
the CCFR experiment at Fermilab,
the NOMAD experiment at CERN, and
the KARMEN experiment at ISIS.
}
\label{fig:Fig{27}_lhd}
\end{figure*}
%
%
\begin{figure*}[\figoptions]
\vspace*{\fill}
\centering
\scalebox{\figscale}{
\includegraphics{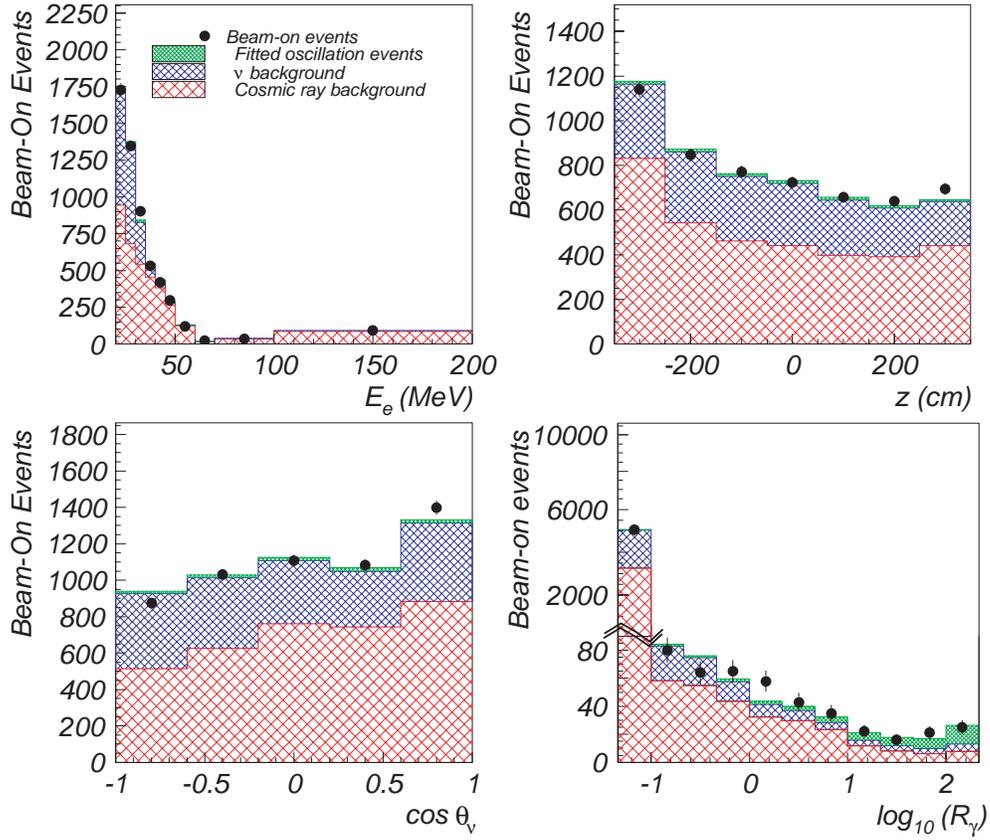}
}
\vspace{5mm}
\caption{
The  $E_e$,  {\it z}, $\cos\theta_\nu$, and $R_\gamma$ projections 
from the 4-dimensional
$(\sin^22\theta,\Delta m^2)$ likelihood fit. The points with error bars are
the data.
}
\label{fig:Fig{28}_projections}
\end{figure*}

%
%
%
\clearpage

%
%
\end{document}